\documentclass[options]{JHEP3}

\usepackage{epsf,epsfig}
\usepackage{amssymb}
\usepackage{graphicx}
\usepackage{amsmath}
\usepackage{amsfonts}
 
\usepackage{graphicx}

\newcommand{\insertplot}[5]{\begin{figure}
 \hfill\hbox to 0.05in{\vbox to #5in{\vfill
 \inputplot{#1}{#4}{#5}}\hfill}
 \hfill\vspace{-.1in}
 \caption{#2}\label{#3}
 \end{figure}}
 \newcommand{\inputplot}[3]{
 \special{ps: plotfile #1}
}
\newcounter{fig}

\newcommand{\beq}{\begin{equation}}
\newcommand{\eeq}{\end{equation}}
\newcommand{\beqs}{\begin{eqnarray}}
\newcommand{\eeqs}{\end{eqnarray}}

\newcommand{\be}{\begin{equation}}
\newcommand{\ee}{\end{equation}}
\newcommand{\bea}{\begin{eqnarray}}
\newcommand{\eea}{\end{eqnarray}}

\numberwithin{equation}{section}

\abstract{ 
We construct rotating black hole solutions  
in Einstein-Gauss-Bonnet theory in five spacetime dimensions.
These black holes are asymptotically flat,  
and possess a regular horizon of spherical topology and 
two equal-magnitude angular momenta 
associated with two distinct planes of rotation.
The action and global charges of the solutions are
obtained by using the quasilocal formalism 
with boundary counterterms generalized 
for the case of Einstein-Gauss-Bonnet theory. 
We discuss the general properties of these black holes 
and study their dependence on the Gauss-Bonnet coupling constant $\alpha$.
We argue that most of the properties of the configurations
are not affected by the higher derivative terms.
For fixed $\alpha$ the set of black hole solutions 
terminates at an extremal black hole with a regular horizon,
where the Hawking temperature vanishes
and the angular momenta attain their extremal values.
The domain of existence of regular black hole solutions is studied.
 The near horizon geometry of the extremal solutions is
determined
by employing the entropy function formalism.
}

\keywords{Einstein-Gauss-Bonnet theory, black holes, numerical solutions}\preprint{ }

\title{Rotating black holes with equal-magnitude angular momenta in $d=5$ Einstein-Gauss-Bonnet theory } 
 
\author{{\large Yves Brihaye}$^{\dagger}$, 
{\large Burkhard Kleihaus}$^{\ddagger}$, {\large Jutta Kunz}$^{\ddagger}$ 
and {\large Eugen Radu}$^{\star \diamond }$ \\ \\
$^{\dagger}${\small Physique-Math\'ematique, Universit\'e de
Mons-Hainaut, Mons, Belgium}\\
$^{\ddagger}${\small Institut f\"ur Physik, Universit\"at Oldenburg, Postfach 2503
D-26111 Oldenburg, Germany}
  \\
$^{\star}${\small  Department of Computer Science,
National University of Ireland Maynooth,
Maynooth,
Ireland} \\
$^{\diamond}${\small School of Theoretical Physics -- DIAS, 10 Burlington
Road, Dublin 4, Ireland }
  }

\begin{document}

\section{Introduction}

In $d\geq 5$ dimensions, the gravity action may be modified to include higher order curvature terms while keeping
the equations of motion to second order, 
provided the higher order terms appear in specific combinations \cite{Lovelock:1971yv}.
In five dimensions, this leads to the so-called 
Einstein-Gauss-Bonnet (EGB) theory, which contains quadratic powers of 
the curvature.
The  Gauss-Bonnet (GB) term appears  in the one-loop corrected
effective action of heterotic string theory~\cite{Gross:1986iv,Myers:1987yn}, 
and also in the low-energy effective action
for the compactification of M-theory on a Calabi-Yau threefold  \cite{Antoniadis:1997eg}.
Inclusion of this term in the action leads to a variety of 
new features (see the recent reviews \cite{Garraffo:2008hu,Charmousis:2008kc}).
In particular, the  black holes of EGB theory do not in general obey the
Bekenstein-Hawking area law,
but the entropy formula includes a new
contribution coming from the higher curvature terms in the action 
\cite{Jacobson:1993xs,Wald:1993nt}.
Also, for $d=5$, the GB term implies the existence of 
a branch of small static black holes which are thermodynamically stable
\cite{Myers:1988ze}.

However, the complexity of EGB theory  makes the task of finding
closed form solutions a highly non-trivial task. The only known
analytic solutions correspond to the counterparts of the 
Schwarzschild-Tangherlini black holes \cite{Boulware:1985wk,Wheeler:1985nh},
and a variety of physically interesting solutions were found only numerically,
  ($e.g.$ the black strings  \cite{Kobayashi:2004hq,Brihaye:2010me}
and the  black rings  \cite{Kleihaus:2009dm}).
In particular, no rotating EGB solutions are known yet
in closed form, and it was proven in \cite{Anabalon:2009kq} that the
Kerr-Schild ansatz may not work in Lovelock theory.
Nevertheless, a number of partial results support the idea that
EGB generalizations of the Myers-Perry (MP) solutions
\cite{Myers:1986un} actually exist \cite{Kim:2007iw,Alexeyev:2007sd}.
 
The main purpose of this work is to present numerical evidence
for the existence of a class of asymptotically flat, rotating solutions
in $d = 4 + 1$ EGB theory.
These solutions have a spherical horizon topology
and two equal-magnitude angular momenta.
This restriction leads to a system of coupled nonlinear ordinary
differential equations, which are solved numerically  
within a nonperturbative approach.
The same approach has been employed to construct
Einstein-Maxwell  \cite{Kunz:2005nm,Kunz:2006eh},
Einstein-Maxwell-Chern-Simons \cite{Kunz:2005ei},
and Einstein-Yang-Mills \cite{Brihaye:2007tw}
rotating black hole solutions in higher dimensions.
Also, the existence of the $AdS$ counterparts of the 
solutions addressed in this work was established in \cite{Brihaye:2008kh}. 
 
Our results suggest that the  EGB rotating black holes
with two equal angular momenta
share most of the features of the corresponding Einstein gravity MP solutions.
In particular, for a fixed value of the GB coupling constant, 
the sets of black hole solutions terminate 
when they reach an extremal black hole, where
the Hawking temperature vanishes and
the equal-magnitude angular momenta assume their extremal values. 
  These extremal black holes obey the attractor mechanism assuming their
near horizon geometry involves an $AdS_2$ factor.
As a new feature, a GB term in the action leads to the existence of
a branch of rotating black holes with a
positive
specific heat at constant angular velocity at the horizon.

This paper is organized as follows:
in the next Section we recall the EGB action and present the ansatz
and boundary conditions.
Section 3 contains a  discussion of the counterterm approach
for asymptotically flat solutions in EGB theory,
together with the computation 
of the action and global charges for rotating black holes with equal magnitude angular momenta.
We present our numerical results in Section 3, and 
exhibit the physical properties of these solutions
and their domain of existence. 
The near horizon geometry of the extremal solutions is also studied 
there by using the entropy function formalism
\cite{Astefanesei:2006dd}. 
We conclude in Section 4
with some further remarks.

\section{The model} 
\subsection{The action  } 
We consider the Einstein-Hilbert action supplemented by the GB term:
\begin{eqnarray}
\label{action}
I=\frac{1}{16 \pi G}\int_{\cal M}~d^5x \sqrt{-g} \left(R +\frac{\alpha}{4}L_{GB}\right),
\end{eqnarray}
where $R$ is the Ricci scalar and 
\begin{eqnarray}
\label{LGB}
L_{GB}=R^2-4R_{\mu \nu}R^{\mu \nu}+R_{\mu \nu \sigma \tau}R^{\mu \nu \sigma \tau}
\end{eqnarray}
is the GB term.
Variation of the action (\ref{action}) with respect to the metric
tensor results in the equations of EGB gravity
\begin{eqnarray}
\label{eqs}
R_{\mu \nu } -\frac{1}{2}Rg_{\mu \nu} +\frac{\alpha}{4}H_{\mu \nu}=0~,
\end{eqnarray}
where
\begin{equation}
\label{eq1}
H_{\mu \nu}=2(R_{\mu \sigma \kappa \tau }R_{\nu }^{\phantom{\nu}%
\sigma \kappa \tau }-2R_{\mu \rho \nu \sigma }R^{\rho \sigma }-2R_{\mu
\sigma }R_{\phantom{\sigma}\nu }^{\sigma }+RR_{\mu \nu })-\frac{1}{2}%
L_{GB}g_{\mu \nu }  ~.
\end{equation}
For a well-defined variational principle, one has to supplement the 
action (\ref{action}) with the Gibbons-Hawking surface term 
\begin{equation}
I_{b}^{(E)}=-\frac{1}{8\pi G}\int_{\partial \mathcal{M}}d^{4}x\sqrt{-h }K~,
\label{Ib1}
\end{equation}
and its counterpart for EGB gravity  \cite{Myers:1987yn} 
\begin{equation}
I_{b}^{(GB)}=-\frac{\alpha}{16\pi G}\int_{\partial \mathcal{M}}d^{4}x\sqrt{-h }%
 \left( {\cal J}-2{\rm G}_{ab} K^{ab}\right)  ~,
\label{Ib2}
\end{equation}
where $h _{ab }$ is the induced metric on the boundary,  $K$ is  the
trace of the extrinsic curvature of the boundary,
  ${\rm G}_{ab}$ is the Einstein tensor of the metric $h _{ab}$ and ${\cal J}$ is the
trace of the tensor
\begin{equation}
{\cal J}_{ab}=\frac{1}{3}%
(2KK_{ac}K_{b}^{c}+K_{cd}K^{cd}K_{ab}-2K_{ac}K^{cd}K_{db}-K^{2}K_{ab})~.
\label{Jab}
\end{equation}
An interesting feature of EGB gravity is the presence 
of two branches of static solutions, distinguished by their behaviour
for $\alpha \to 0$ \cite{Boulware:1985wk}. 
In this paper we shall restrict our analysis of rotating
EGB black hole solutions to those, whose static limit
corresponds to the branch of static solutions 
with a well defined Einstein gavity limit.

 \subsection{The ansatz and boundary conditions}  

While rotating EGB black holes will generically possess two independent
angular momenta and a more general topology 
of the event horizon\footnote{Recently, also  static  black rings were obtained
in $d=5$ EGB gravity by employing a numerical approach \cite{Kleihaus:2009dm}.}
we restrict here to configurations with 
equal-magnitude angular momenta and a spherical horizon topology.
A suitable metric ansatz reads  \cite{Kunz:2005nm}
\begin{eqnarray}
\label{metric}
&&ds^2 = \frac{dr^2}{f(r)}
  + g(r) d\theta^2
+h(r)\sin^2\theta \left( d \varphi_1 -w(r)dt \right)^2 
+h(r)\cos^2\theta \left( d \varphi_2 -w(r)dt \right)^2 ~~{~~~~~}
\\
\nonumber
&&{~~~~~~}+(g(r)-h(r))\sin^2\theta \cos^2\theta(d \varphi_1 -d \varphi_2)^2
-b(r) dt^2,
\end{eqnarray}
where $\theta  \in [0,\pi/2]$, $(\varphi_1,\varphi_2) \in [0,2\pi]$, 
and $r$ and $t$ denote the radial and time coordinate, respectively.
  Also, this metric ansatz admits a simpler expression in terms of
the left-invariant
1-forms $\sigma_i$ on $S^3$, with 
\begin{eqnarray}
\label{metric2}
&&ds^2 = \frac{dr^2}{f(r)}
  + \frac{1}{4}g(r)(\sigma_1^2+\sigma_2^2)+\frac{1}{4}h(r) \big(\sigma_3+2w(r) dt \big)^2
-b(r) dt^2,
\end{eqnarray}
where 
$\sigma_1=\cos \psi d\bar \theta+\sin\psi \sin  \theta d \phi$, 
$\sigma_2=-\sin \psi d\bar \theta+\cos\psi \sin  \theta d \phi$,
$\sigma_3=d\psi  + \cos  \theta d \phi$ 
and $2\theta=\bar \theta$,
 $\phi_1-\phi_2=\phi$,
  $\phi_1+\phi_2=\psi$.

For such solutions the isometry group is enhanced from $R \times U(1)^{2}$
to $R \times U(2)$, where $R$ denotes the time translation.
  This symmetry enhancement allows to factorize the angular dependence 
and thus leads to ordinary differential equations.
 
Without fixing a metric gauge, a straightforward computation
leads to the following reduced action for the system
 \begin{eqnarray}
\label{Leff} 
A_{\rm eff}=\int dr dt ~L_{\rm eff},~~~{\rm with~~~~} 
L_{\rm eff}=L_E+\frac{\alpha}{4}L_{GB},
\end{eqnarray}
with
\begin{eqnarray}
L_{E}&=&
\sqrt{\frac{fh}{b}}
\bigg(
b'g'+\frac{g}{2h}b'h'+\frac{b}{2g}g'^2+\frac{b}{h}g'h'+\frac{1}{2}gh w'^2+\frac{2b}{f}(4-\frac{h}{g})
\bigg),
\\
\nonumber
 L_{GB}&=&\sqrt{\frac{fh}{b}}\frac{1}{ g}
\left(
\frac{4h}{g}b'g'+2(4g-3h)(\frac{b'h'}{h}+hw'^2)
-\frac{f}{2h}b'h'g'^2-\frac{1}{2}fhg'^2w'^2
\right),
\end{eqnarray}
 where a prime denotes a
derivative with respect to $r$.
The corresponding equations for the metric functions 
$b$, $f$, $h$ and $w$ are found by taking the variation
of $A_{\rm eff}$ with respect to $a$, $b$, $f$ and $g$ 
and by fixing afterwards the metric gauge. 
(This procedure is equivalent to solving the EGB equations directly, 
but it is technically simpler.)
 The equation for $w(r)$ admits the first integral
\begin{eqnarray}
\label{FIw}
w'h\sqrt{\frac{fh}{b}}(g-\frac{\alpha}{4}(\frac{fg'^2}{g}-16+\frac{12h}{g}))=const.,
\end{eqnarray}
which is useful in the numerical calculations.
The remaining equations are rather long, and we do not include them here. 
In the following\footnote{The
numerical calculations are performed in two gauges.
 Besides the gauge $g(r)=r^2$, we also employ the gauge
choice in \cite{Kunz:2005nm,Kunz:2006eh} 
(with $b(r)=\bar f(r),~f=\bar f(r)/m(r),~g(r)=r^2 m/\bar f(r),~ h(r)=n r^2/\bar f(r)$,
and $w(r)=\bar w(r)/r$),
corresponding to isotropic coordinates.
}  we fix the metric gauge by taking $g(r)=r^2$.

Also, the EGB equations (\ref{eqs}) imply the following relations 
which will be important in the later discussion
\begin{eqnarray}
\label{totder1} 
&&\frac{1}{\sin^2 \theta}(R_\varphi^t+\frac{\alpha}{4}H_\varphi^t)
=\frac{1}{\cos^2 \theta}(R_\psi^t+\frac{\alpha}{4}H_\psi^t)
\\
\nonumber
&&
{~~~~~~~~~~~~~~~~~~~~~~~~~}=  \frac{1}{2 r^2} \sqrt{\frac{f}{ b h}}\frac{d }{dr}  
 \left [\sqrt{\frac{f h}{b}} h w'
 \bigg (
- r^2 +  \alpha ( f-4+ \frac{3h}{r^2})\bigg)
\right],
\\ 
\label{totder2} 
&&R_t^t+\frac{\alpha}{4}(H_t^t+\frac{1}{2}L_{GB})=
\frac{1}{2r^2} \sqrt{\frac{f}{ b h}}  \frac{d }{dr}
 \bigg [\sqrt{\frac{f h}{b}}
 \bigg( r^2(hww'-b')
 \\
 \nonumber
&&
{~~~~~~~~~~~~~~~~~~~~~~~~~~~~~~~}
+\alpha\left((f-4+\frac{h}{r^2}+\frac{rfh'}{h})b'
 +(4-\frac{3h}{r^2}-f)hww'+rfhw'^2\right)
  \bigg )
\bigg ] .
\end{eqnarray} 


The horizon of these black hole solutions is a squashed $S^3$ sphere. 
It resides at the constant value of
the radial coordinate $r=r_H>0$, and is characterized by $f(r_H)=b(r_H) =  0$. 
Restricting to nonextremal solutions, the following expansion
holds near the event horizon: 
\begin{eqnarray}
\label{c1}
&&f(r)=f_1(r-r_H)+  O(r-r_H)^2,~~h(r)=h_H+ O(r-r_H),
\\
\nonumber
&&
b(r)=b_1(r-r_H)+O(r-r_H)^2,~~w(r)=\Omega_H+w_1(r-r_H)+O(r-r_H)^2.~{~ }
\end{eqnarray}  
For a given event horizon radius, 
the essential parameters characterizing the event horizon
are $f_1,~b_1$,~$h_H$,~$\Omega_H$ and $w_1$ (with $f_1>0,~b_1>0$),
which fix all higher order coefficients in (\ref{c1}).
(These constants are related in a complicated way to the global
charges of the solutions.)
The (constant) horizon angular velocity $\Omega_H$
is defined in terms of the Killing vector
$\chi=\partial/\partial_t+
 \Omega_1 \partial/\partial \varphi_1 + \Omega_2 \partial/\partial \varphi_2 $
which is null at the horizon.
For the solutions within the ansatz (\ref{metric}), the 
horizon angular velocities are equal, $\Omega_1=\Omega_2=\Omega_H$.

A straightforward calculation gives the following asymptotic expansion 
for the metric functions,
involving three parameters ${\cal U},~{\cal V}$ and ${\cal W}$
\begin{eqnarray}
\nonumber
&&b(r)=
1
+\frac{{\cal U}}{r^2}
+\left(2{\cal W}^2-{\cal U}{\cal V}+3 {\cal U}^2\frac{\alpha}{2} \right)\frac{1}{3r^6}
+\left({\cal U}^2{\cal V}+ {\cal U}-2{\cal W}^2{\alpha}\right )\frac{1}{3r^8}+O\big(\frac{1}{r^{10}}\big),
\\
\nonumber
&&f(r)=
1
+\frac{{\cal U}}{r^2}
+\frac{{\cal V}}{r^4}
+\left(-{\cal W}^2-{\cal U}{\cal V}+ {\cal U}^2\frac{\alpha}{2}\right)\frac{1}{r^6}
\\
\nonumber
&&{~~~~~~~~~~~~~~~~~~~~~~~~~~~~~~~~~~}
+\left(\frac{23}{30}{\cal U}({\cal W}^2+{\cal U}{\cal V})+\frac{14}{15}({\cal W}^2-3{\cal U}{\cal V}) {\alpha}\right)\frac{1}{r^8}+O\big(\frac{1}{r^{10}}\big),
\\
\label{inf1}
&&h(r)=r^2 
+\frac{{\cal V}}{r^2}
-\frac{{\cal W}^2+{\cal U}{\cal V}}{r^4}
+\left(\frac{9{\cal U}}{10}({\cal W}^2+{\cal U}{\cal V})+\frac{2}{5}({\cal W}^2-3{\cal U}{\cal V})\alpha\right)\frac{1}{r^6}
\\
\nonumber
&&{~~~~~~~~~~~~~~~~~~~~~~~~~~~~~~~~~~}
+\frac{1}{15}\bigg( (5{\cal V}-12{\cal U}^2)({\cal W}^2+{\cal U}{\cal V})-7{\cal U} ({\cal W}^2-3{\cal U}{\cal V})\alpha \bigg)\frac{1}{r^8}
+O\big(\frac{1}{r^{10}}\big),
\\
&&
\nonumber
w(r)=\frac{{\cal W}}{r^4}
-\frac{{\cal W}({\cal V}-{\cal U}\frac{\alpha}{2})}{r^8}+O\big(\frac{1}{r^{10}}\big),
\end{eqnarray}
which guarantees that the Minkowski spacetime background 
is approached at infinity.

\subsection{Known solutions and slowly rotating black holes}

For the metric ansatz (\ref{metric}), the EGB field equations (\ref{eqs})  
possess two well known exact solutions. 
First, the MP black holes \cite{Myers:1986un}  
with equal-magnitude angular momenta
are found for $\alpha=0$ (i.e., without GB term).
Expressed in terms of the event horizon radius 
and the horizon angular velocity\footnote{This metric is usually expressed
in terms of the mass parameters $M$ and the angular momentum parameter $a$, 
with $M=r_H^2/2(1-r_H^2\Omega_H^2)$ and $a=r_H^2\Omega_H$.}
(which are the control parameters in our numerical approach), 
this solution reads
\begin{eqnarray}
\nonumber
 &&f(r)=1-\frac{1}{1-r_H^2\Omega^2_H}\left(\frac{r_H}{r}\right)^2
 +\frac{r_H^2\Omega_H^2}{1-r_H^2\Omega_H^2}\left(\frac{r_H}{r}\right)^4,
~~
 h(r)=r^2
 \left(
 1+\left(\frac{r_H}{r}\right)^4\frac{r_H^2\Omega_H^2}{1-r_H^2\Omega_H^2}
  \right),
\\
\label{MP}
&&b(r)=1
 -\left(\frac{r_H}{r}\right)^2
 \frac{1}{1-\left(1-(\frac{r_H}{r})^4\right)r_H^2\Omega_H^2},
 ~~w(r)=(\frac{r_H}{r})^4\frac{ \Omega_H}{1-\left(1-(\frac{r_H}{r})^4\right)r_H^2\Omega_H^2}.
\end{eqnarray}
Therefore, for a MP black hole, the relevant parameters 
in the event horizon expansion (\ref{c1}) are
\begin{eqnarray}
\label{g1}
f_1=\frac{2 (1-2r_H^2\Omega_H^2)}{r_H(1-r_H^2\Omega_H^2)},
 ~~
b_1= \frac{2}{r_H}(1-2r_H^2\Omega_H^2),
 ~~
w_1= \frac{4\Omega_H}{r_H}( r_H^2\Omega_H^2-1),
\end{eqnarray}
while the constants ${\cal U},~{\cal V}$ and ${\cal W}$ 
in the far field expansion (\ref{inf1}) have the following 
expression\footnote{Note that since ${\cal V}=-{\cal W}^2/{\cal U}$, 
there are only two free parameters in the far field expansion,
which fix the mass and equal-magnitude angular momenta of the solutions.} 
\begin{eqnarray}
\label{g2}
 {\cal V}=\frac{r_H^6 \Omega_H^2}{1-r_H^2\Omega_H^2},
 ~~
  {\cal U}=-\frac{r_H^2  }{1-r_H^2\Omega_H^2},
 ~~
  {\cal W}= \frac{r_H^4 \Omega_H }{1-r_H^2\Omega_H^2}.
\end{eqnarray}

The $d=5$ MP black holes with equal-magnitude angular momenta 
emerge smoothly from the static Schwarzschild-Tangherlini black hole 
when the event horizon velocity $\Omega_H$ is increased from zero. 
For a given event horizon radius, 
the solutions exist up to a maximal value of 
the horizon angular velocity $\Omega_H^{(c)}=1/\sqrt{2}r_H$.
Expressed in terms of the mass-energy $E$ and the equal-magnitude
angular momenta $|J_1|=|J_2|=J$, 
this bound reads $27 \pi J^2/8G<E^2$.
The extremal solution saturating this bound 
has a regular but degenerate horizon.
Further details on the properties of MP black holes 
with equal-magnitude angular momenta are found in 
\cite{Kleihaus:2007kc,Brihaye:2009kz}.

The second exact solution corresponds to the generalization 
\cite{Boulware:1985wk} of the static 
Schwarzschild solution with a GB term and reads
(note our restriction to the branch with Einstein gravity limit)
\begin{eqnarray}
\label{SGB}
 w(r)=0,~~ h(r)=r^2~~{\rm and}~~
 f(r)=b(r)=1+\frac{r^2}{\alpha}\left(1-\sqrt{1+\frac{\alpha}{r^4}(\alpha+2r_H^2)}\right).
\end{eqnarray} 
This solution exists for all $r_H > 0$
and $\alpha>-r_H^2$ (with $r_H$ the event horizon radius).
Without entering into details, 
let us mention the existence of some substantial differences between
the thermodynamics of the static $d = 5$ EGB black hole solutions\footnote{For $d>5$,
the thermodynamical properties of the solutions are similar to the Einstein gravity case.}
and their Einstein gravity counterparts 
(restricting to the physical case $\alpha> 0$).
If the black holes are large enough, $r_H^2>\alpha $, 
then they behave like their Einstein gravity counterparts since the
specific heat is negative, $C_p=T_H(\partial S/\partial T_H)< 0$. 
A different behaviour is found for $r_H^2/\alpha<1$, 
since in that case $T_H\simeq r_H$ and thus $C_p > 0$.  
This implies the existence of a branch of small five-dimensional 
EGB black holes that is thermodynamically stable \cite{Myers:1988ze}
(see $e.g.$~\cite{Garraffo:2008hu} for a review of these aspects).

{To the best of our knowledge, 
there is no exact solution of the EGB equations
describing non-static configurations\footnote{See however, the results in \cite{Anabalon:2009kq}.
However, the solution there is very special ($e.g.$ the value of the GB coupling constant is fixed by the cosmological
constant)
and does 
not describe a black object.}.}
However, slowly rotating 
black holes can be found by considering perturbation theory 
around the static solution (\ref{SGB}) in terms 
of the rotation parameter $a$ (see $e.g.$~\cite{Kim:2007iw}).
For our case 
the slowly rotating solution then contains in its non-diagonal metric elements
the function $w(r)$
\begin{equation}
\label{w-slow}
w(r)=\frac{a(2 r_H^2+\alpha)}{ r^4 \left(1+\sqrt{1+\frac{\alpha (2r_H^2 +\alpha)}{r^4}}\right)} ,
\end{equation} 
that is linear in the perturbative parameter $a$,
while the other metric functions remain unchanged 
to this order in $a$.


\section{The boundary counterterm approach in
$d=5$ Einstein-Gauss-Bonnet theory and the global charges}
\subsection{A counterterm for the $d=5$ asymptotically flat space }

It is well known that the gravitational action 
contains divergences even at the tree-level --
that arise from integrating over the infinite volume of spacetime.
A common approach - background subtraction - uses a second, reference spacetime
to identify divergences which should be subtracted from the action. 
After subtracting the (divergent) action of the reference background, 
the resulting action will be finite.

At a conceptual level, the background subtraction method is 
not entirely satisfactory, since it relies on
the introduction of a spacetime which is auxiliary to the problem. 
In some cases the choice of reference
spacetime is ambiguous - for example NUT-charged solutions 
(see $e.g.$~the discussion in \cite{Mann:2002fg}).
For asymptotically $AdS$ spacetimes, 
this problem is solved by adding additional surface terms to the
action \cite{Balasubramanian:1999re}. 
These counterterms are built up with curvature invariants 
of a boundary ${\partial {\cal M}}$ 
(which is sent to infinity after the integration),
and thus they do not alter the bulk equations of motion.
This yields a finite action and mass of the system. 

The generalization of this procedure to the asymptotically flat case was
considered in \cite{Mann:1999pc,Lau:1999dp,Kraus:1999di}. 
Moreover, as discussed  in \cite{Astefanesei:2005ad},
a renormalized stress-tensor can be defined by
varying the total action (including the counterterms) 
with respect to the boundary metric. 
The conserved quantities can be constructed from this stress tensor 
via the algorithm of Brown and York \cite{Brown:1993br}.

However, all studies in the literature of asymptotically flat configurations
have considered the case of counterterms in Einstein gravity 
only\footnote{The general 
expression of the counterterms and the boundary stress tensor 
in EGB theory with a cosmological constant is presented in 
\cite{Brihaye:2008xu,Brihaye:2008ns}.}.
We have found that for five-dimensional 
asymptotically flat EGB  solutions with a boundary topology $S^3 \times R$, 
the action can be regularized
by the following counterterm:
\begin{eqnarray}
\label{Ict}
I_{ct}=-\frac{1}{8 \pi G}\int_{\partial {\cal M}} d^{4} x \sqrt{-h}\Psi(\mathcal{R}),
\end{eqnarray}
where $\mathcal{R}$ is the Ricci scalar of the induced metric on the
boundary $h_{ij}$
and
\begin{eqnarray}
\label{Ict-n}
\Psi(\mathcal{R})=
\sqrt{\frac{3}{2}\,\mathcal{R}}\left(1+\frac{\alpha}{9}\mathcal{R} \right).
\end{eqnarray}
For $\alpha=0$, this reduces to the counterterm expression proposed in 
\cite{Kraus:1999di}
for Einstein gravity with the same asymptotic 
structure. 

Varying the total action (which contains the Gibbons-Hawking boundary term)
with respect to the
boundary metric $h_{ij}$, we obtain the divergence-free boundary stress-tensor
\begin{eqnarray}
\label{Tik}
&T_{ab}=\frac{2}{\sqrt{-h}}\frac{\delta I}{\delta h^{ab}}=
\frac{1}{8\pi G}\Big( K_{ab}-h_{ab}K
+ \frac{{\alpha}}{2} (Q_{ab}-\frac{1}{3}Qh_{ab})
-2\frac{d\Psi}{d \mathcal{R}}\mathcal{R}_{ab}
+2\Psi h_{ab} 
-2h_{ab}\Box \Psi
+\Psi_{;ab}
\Big),~~{~~~}
\end{eqnarray} 
where $K_{ab}$ is the extrinsic curvature of the boundary 
and \cite{Davis:2002gn}
\begin{eqnarray}
\nonumber
Q_{ab}= 
2KK_{ac}K^c_b-2 K_{ac}K^{cd}K_{db}+K_{ab}(K_{cd}K^{cd}-K^2)
+2K \mathcal{R}_{ab}+\mathcal{R}K_{ab}
-2K^{cd}\mathcal{ R}_{cadb}-4 \mathcal{R}_{ac}K^c_b~,
\end{eqnarray}
while $\mathcal{R}_{abcd}$, $\mathcal{R}_{ab}$  denote
the Riemann   and Ricci tensors of the boundary metric.

Provided the boundary geometry has an isometry generated by a
Killing vector $\xi ^{i}$, a conserved charge
\begin{eqnarray}
{\frak Q}_{\xi }=\oint_{\Sigma }d^{3}S^{i}~\xi^{j}T_{ij}
\label{charge}
\end{eqnarray}
can be associated with a closed surface $\Sigma $ \cite{Mann:2002fg}. 
Physically this means that a collection of observers on
the hypersurface with metric $h_{ij}$ all observe the same value
of ${\frak Q}_{\xi }$, provided this surface has an isometry
generated by $\xi$.  

 To test the counterterm (\ref{Ict}), we have verified
that the known expressions for the mass and action are recovered
 for the case of the Schwarzschild-Tangherlini solution in EGB.
 Neglecting the second term
in (\ref{Ict-n}) ($i.e.$ considering only the Einstein gravity counterterm) 
results in the occurrence of an unphysical constant term  
proportional with $\alpha$  in the expressions of $I_{cl}$ and $E$.
 
 However, one should remark that the proposal (\ref{Ict}) inherits all ambiguities already present in the 
 Einstein gravity case.
 In particular, there is no rigorous justification for considering that expression,
 and, in fact
the  results in \cite{Kraus:1999di}  show that the counterterm choice in Einstein gravity 
is not unique.
 This suggests that other more complicated expressions are possible for the EGB case 
 as well\footnote{However, the expressions for the regularized action and the 
 conserved quantities will be the same for any counterterm choice.}.
 Moreover,  unlike the $AdS$ case, the counterterm  depends  on the
boundary topology  
(for example, we have verified that the coefficient 
in front of (\ref{Ict}) is different for black string solutions).

These problems are avoided in the proposal put forward by Mann and Marolf
in \cite{Mann:2005yr}
 for asymptotically flat spacetimes
and further generalized in \cite{Kleihaus:2009ff}
to extended objects, like black strings and $p-$brane spacetimes.
In this approach, the
conserved quantities are constructed essentially using the electric part
of the Weyl conformal tensor.
The generalization of the  Mann and Marolf prescription to Lovelock gravity solutions
is an interesting open problem.

\subsection{The action and global charges of solutions}
The computation of the boundary stress-tensor
$T_{ab}$   for the solutions of interest  in this work is straightforward, and, for the asymptotic
expression (\ref{inf1}), we find the 
following expressions for the  relevant components
\begin{eqnarray}
\label{tik} 
 T_{\varphi_{_1}}^t=\frac{1}{8\pi G}\frac{2{\cal W}\sin^2 \theta}{r^3}+O(1/r^5),~
 T_{\varphi_{_2}}^t=\frac{1}{8\pi G}\frac{2{\cal W}\cos^2 \theta}{r^3}+O(1/r^5),~ 
 T_t^t=\frac{1}{8\pi G}\frac{3{\cal U} }{2r^3}+O(1/r^5).~~{~~}
\end{eqnarray}
The energy $E$ and the equal-magnitude angular momenta $J$ are the
charges associated with the Killing vectors $\partial/\partial t$, 
$\partial/\partial \varphi_1$, and $\partial/\partial \varphi_2$, 
respectively. 
Computed according to (\ref{charge})  these quantities are 
\begin{eqnarray}
\label{global-charges}
E=-\frac{3V_3 }{16\pi G}{\cal U},~~
J=\frac{ V_3 }{8\pi G}{\cal W},
\end{eqnarray}
where $V_3=2\pi^2$ denotes the area of the unit three-dimensional sphere.

Other quantities of interest are 
the Hawking temperature  $T_H=1/\beta$  
and  the  area $A_H$ of the black hole horizon 
\begin{eqnarray} 
\label{Temp-rot} 
  T_H=\frac{\sqrt{b_1f_1}}{4\pi},~~A_H= \sqrt{h_H}  r_H^2 V_3.
\end{eqnarray} 
The thermodynamics of the EGB black holes can be  formulated via the 
path integral approach \cite{Gibbons:1976ue,Hawking:ig}.
Following the standard prescription, 
one computes the classical bulk action evaluated on the equations of motion,
by replacing the $R +\frac{\alpha}{4}L_{GB}$ volume term with
$2(R_t^t+\frac{\alpha}{4}( H_t^t+\frac{1}{2}L_{GB} ))$.
Then one makes use of (\ref{totder2}) to express
the volume integral of this quantity as the difference 
of two boundary integrals.
The boundary integral on the event horizon is simplified by using the identity
(\ref{totder1}) which provides the following relation between the
asymptotic parameter ${\cal W}$ (which fixes the equal-magnitude
angular momenta) and the horizon data which enter (\ref{c1}):
\begin{eqnarray}
{\cal W}=\frac{1}{4}h_H
\sqrt{\frac{f_1 h_H}{b_1}}w_1(-r_H^2+\alpha (\frac{3h_H}{r_H^2}-4)).
\end{eqnarray}
A straightforward calculation using the asymptotic
expressions (\ref{inf1})
shows that the divergencies of the boundary integral at infinity,  
together with the contributions
from $I_{b}^{(E)}$ and $I_{b}^{(GB)}$, are regularized by $I_{\mathrm{ct}}$. 
As a result, 
one finds the following finite expression for the classical action
in terms of the event horizon data and the
asymptotic parameters ${\cal U}$ and ${\cal W}$
\begin{eqnarray}
\label{Icl} 
I_{cl}=\frac{V_3}{4  G}
\bigg(
- 
\left(\sqrt{h_H}r_H^2+\frac{3{\cal U}}{\sqrt{b_1 f_1}}
+\frac{4\Omega_H {\cal W}}{\sqrt{b_1 f_1}}\right)
+\alpha\frac{\sqrt{h_H}}{r_H^2}(h_H -4r_H^2)
\bigg).
\end{eqnarray}
We have verified that a similar result for $I_{cl}$ is obtained 
when using instead the standard 
background subtraction regularization procedure 
(with a Minkowski spacetime background).

Upon application of the Gibbs-Duhem relation to the partition 
function,  one finds  the entropy 
$S=\beta (E- 2\Omega_H J)-I_{cl}$, 
which is the sum of one quarter of the event horizon area
(the Einstein gravity term) plus a GB correction
\begin{eqnarray}
S=S_E+S_{GB},~~{\rm with}~~S_E=\frac{V_3}{4G}r_H^2\sqrt{h_H},~~
S_{GB}=\alpha\frac{ V_3}{4G}\sqrt{h_H}(4-\frac{h_H}{r_H^2}).
\end{eqnarray}
It is interesting to note that the above expression for the entropy  
can also be written in Wald's form \cite{Wald:1993nt}
as an integral over the event horizon 
\begin{eqnarray}
\label{S-Noether} 
S=\frac{1}{4G}\int_{\Sigma_h} d^{3}x \sqrt{\tilde h}(1+\frac{\alpha}{2}\tilde R),
\end{eqnarray} 
(where $\tilde h$ is the determinant of the induced metric on the horizon 
and $\tilde R$ is the event horizon curvature).

\section{The rotating EGB solutions}
\subsection{Non-extremal black holes}

\subsubsection{General features}

In the absence of closed form solutions,
we relied on numerical methods to solve the EGB equations.
In this work\footnote{The numerical methods
here are similar to those used in literature to find numerically  
black hole solutions with equal-magnitude angular momenta in Einstein-Maxwell
theory \cite{Kunz:2005nm,Kunz:2006eh,Kunz:2005ei}.}, we integrated the system of coupled non-linear 
ordinary differential equations 
with appropriate boundary conditions  which follow from (\ref{c1}), (\ref{inf1}),
by using a standard solver \cite{COLSYS}.
This solver involves a Newton-Raphson method for 
boundary-value ordinary differential equations, 
equipped with an adaptive mesh selection procedure.
Typical mesh sizes include $10^2-10^3$ points.
The solutions in this work have a typical relative accuracy of $10^{-6}$.

In our approach, the input parameters\footnote{We have assumed 
that $\Omega_H> 0$, which can always
be achieved by $t \to -t$ if necessary. 
Also, in string theory, the GB coefficient  $\alpha$ is positive,
which is the only case considered here.} 
are the GB coupling constant $\alpha$,
the event horizon radius $r_H$ and the horizon angular velocity
$\Omega_H$ (or equivalently, the equal-magnitude 
angular momenta $J$ through the parameter ${\cal W}$).
Physical quantities characterizing the solutions can then be extracted 
from the numerical solutions.

For most of the analysis, we set $r_H=1$,
which does not spoil the generality of the results,
since it corresponds to fixing the scale of the problem.  
 Also,  to simplify the problem,
 we restrict our integration to the region 
outside the event horizon\footnote{However, similar to 
the static case, the
GB term is expected to drastically affect the geometry in the region inside the horizon
of the rotating black holes (see e.g.~\cite{Torii:2005xu}).}. 

In constructing rotating EGB black holes, 
we made use of the existence of the closed form solutions
(\ref{MP}) and (\ref{SGB}), and employed them as starting configurations,
when increasing gradually $\alpha$ or $\Omega_H$, respectively.
Our results clearly show that 
any such MP solution admits  generalizations in EGB theory.
 When starting instead from the EGB static black holes (\ref{SGB}),
a corresponding branch of rotating EGB solutions 
emerges smoothly for any value of the
event horizon radius $r_H$.

The profiles of the metric functions of  typical EGB black hole solutions
are presented in Figure 1a
for a static ($\Omega_H=0$) and a rotating solution with $\Omega_H=0.68$
(where both solutions have the same value of the GB coupling constant). 
The effects of the GB coupling constant are shown in Figure 1b,
where rotating solutions for two values of $\alpha$ 
and the same horizon angular velocity are shown (here the solution with 
$\alpha=0.01$ is very close to the MP configuration).
One can see that, {apart from a nonzero $w(r)$,}  the rotation leads to non-constant values for $h(r)/r^2$,
while a nonzero $\alpha$ leads to nontrivial deformations 
of all metric functions. 

These rotating EGB solutions possess also an ergoregion 
inside of which the observers cannot remain stationary, 
and will move in the direction of the rotation.
 The ergoregion is bounded by the event horizon, located 
at $r=r_H$ and the stationary limit surface, 
or ergosurface, located at $r=r_e$, where the Killing vector $\partial/\partial t$ 
becomes  null, $i.e.$
$g_{tt}= -b(r_e)+h(r_e) w(r_e)^2=0$.  One can see that the  ergosurface does not intersect the horizon.
 \newpage
\begin{figure}[ht]
\hbox to\linewidth{\hss%
	\resizebox{8cm}{6cm}{\includegraphics{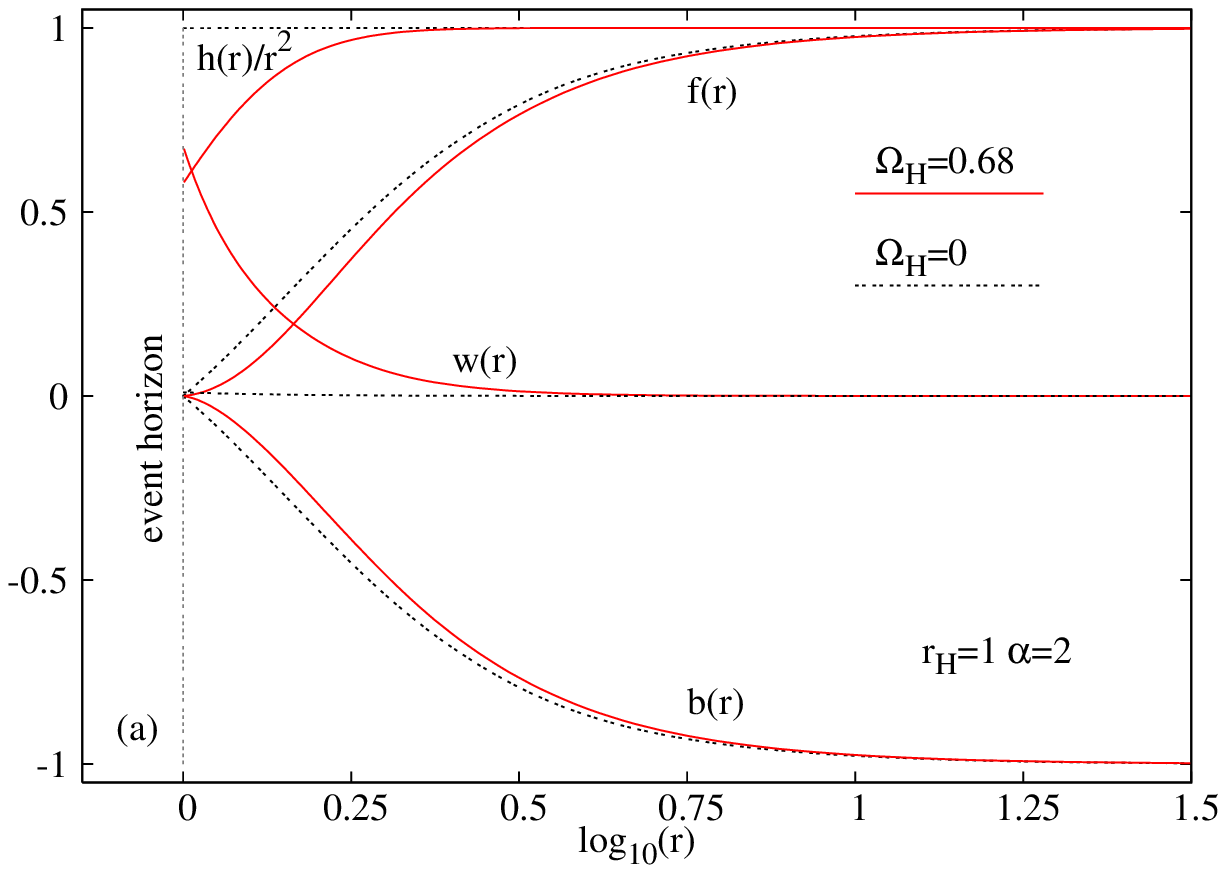}}
\hspace{10mm}%
        \resizebox{8cm}{6cm}{\includegraphics{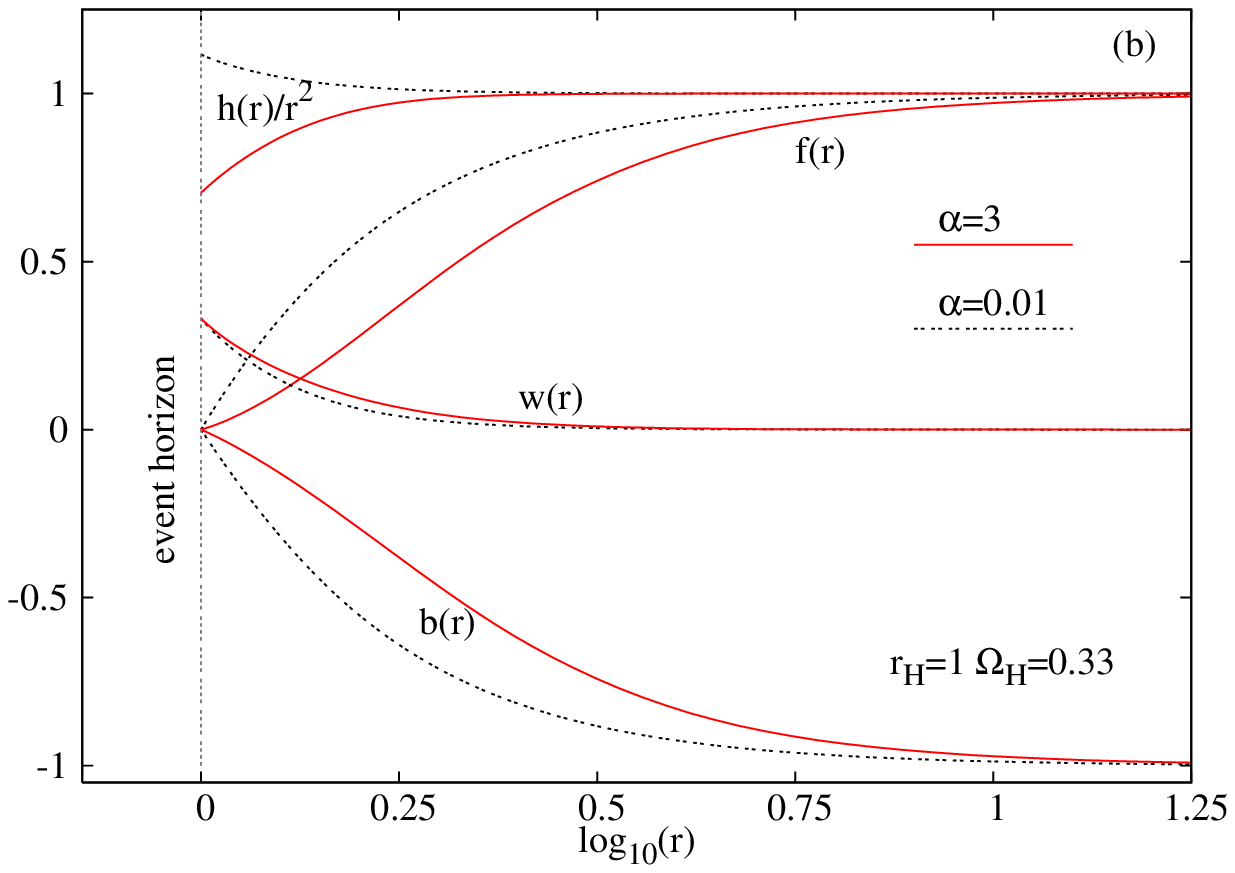}}	
\hss}
\caption{\small 
The metric functions $f(r),~b(r),~w(r)$ and $h(r)/r^2$ 
of typical EGB  black hole solutions  are shown for 
(a) two values of the horizon angular velocity $\Omega_H$
and (b) two values of the Gauss-Bonnet coupling constant $\alpha$.} 
 \end{figure} 

The effect of the GB term is to increase 
the size of the 
ergoregion.
For a fixed event horizon radius $r_H=1$ and $\Omega_H=0.33$ one finds  $e.g.$
$r_e(\alpha=0) \simeq 1.059$, 
$r_e(\alpha=1) \simeq 1.08$
and $r_e(\alpha=2) \simeq 1.104$.

Another interesting feature concerns the sign of the  quantity $\rho_{\rm eff}=\frac{\alpha}{4} H_{t}^t $ 
which, via the modified Einstein equations 
$
G_{\mu\nu}=-\frac{\alpha}{4}  H_{\mu\nu} =T_{\mu\nu}^{\rm eff},
$
corresponds to a
local  `effective energy density'.
This effective stress tensor, 
thought of as a kind of matter distribution,
in principle may violate the weak energy 
condition. Indeed, this is the case for 
both black string \cite{Brihaye:2010me} and black ring \cite{Kleihaus:2009dm}
solutions in EGB theory.  
However, we have found that similar to the case of the static spherically symmetric
black holes (\ref{SGB}), the `effective energy density' is a strictly positive
quantity for all rotating solutions investigated.
This result together with those in \cite{Brihaye:2010me,Kleihaus:2009dm}  
leads to the conjecture that in order to have $\rho_{\rm eff}>0$
the solutions should possess a spherical topology of the event horizon
and be cohomogeneity-1\footnote{ This conjecture could be tested by constructing the $d=5$ caged black holes                  
in $d=5$ EGB theory ($i.e.$ Kaluza-Klein configurations which resemble the Schwarzschild black hole close to the horizon 
and approach asymptotically the four dimensional Minkowski space times a circle).}.

\subsubsection{Domain of existence}

Because EGB black holes are found 
starting both with the MP configurations and with the Schwarzschild-GB static
solutions,
we conclude that, for fixed $r_H$, 
rotating EGB black holes should exist
in a given domain of the $(\alpha, \Omega_H)$ plane.  
To map out this domain, 
we fixed certain values of $\alpha > 0$ 
and increased $\Omega_H$ gradually.
We then obtained numerical solutions up to a maximal value 
of the horizon angular velocity $\Omega_H^{(c)}$, 
that depends on $\alpha$.

The solutions are numerically robust 
but the integration becomes difficult,
when the maximal value $\Omega_H^{(c)}$ is approached.   
However, for any $\alpha > 0$, 
no critical phenomenon (like $e.g.$~a bifurcation
or the approach to a singular point) 
seems to arise there.

In order to clarify the issue of the limiting solutions for
$\Omega_H \to \Omega_H^{(c)}$ for fixed $\alpha$,
the study of the event horizon 
values $b'(r_H)=b_1$ and $f'(r_H)=f_1$ 
as functions of $r_H$ turned
 \newpage
\begin{figure}[ht]
\hbox to\linewidth{\hss%
	\resizebox{8cm}{6cm}{\includegraphics{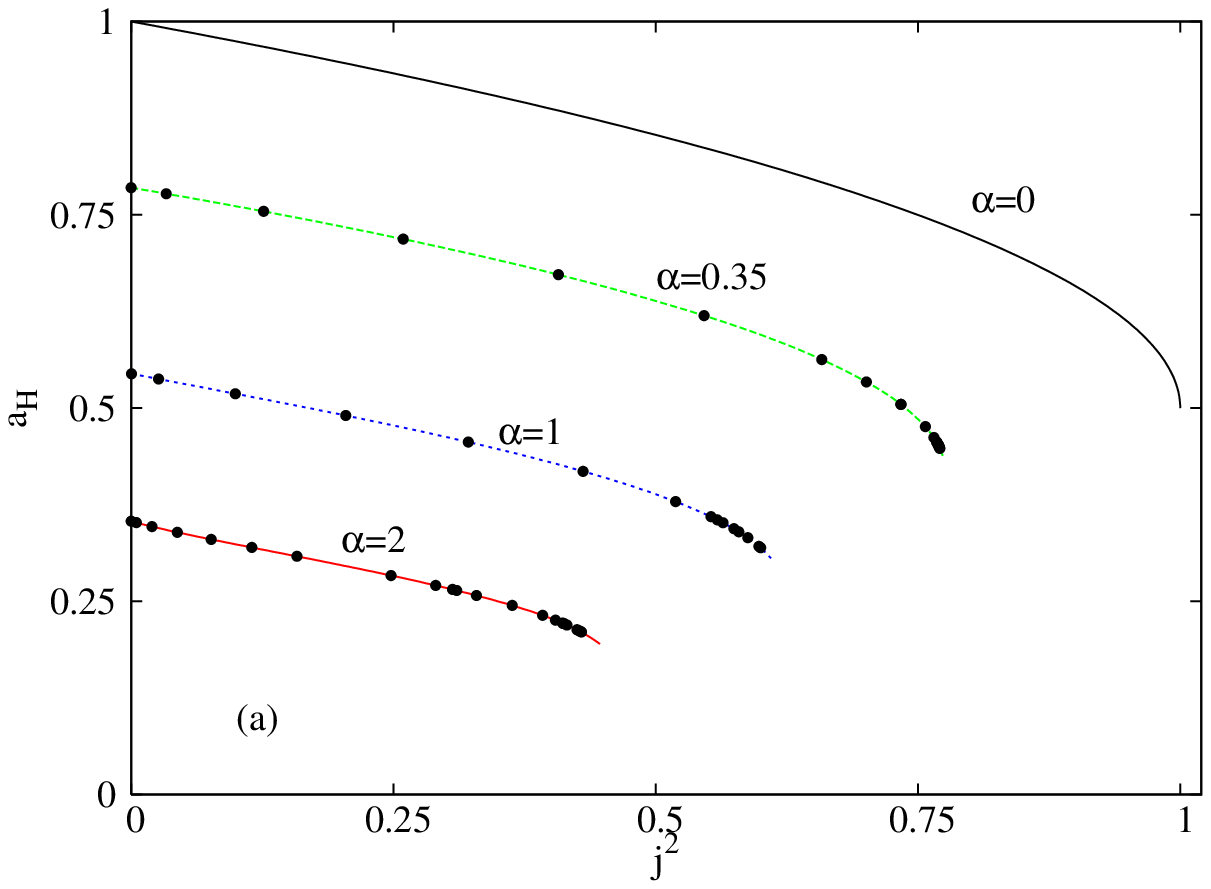}}
\hspace{10mm}%
        \resizebox{8cm}{6cm}{\includegraphics{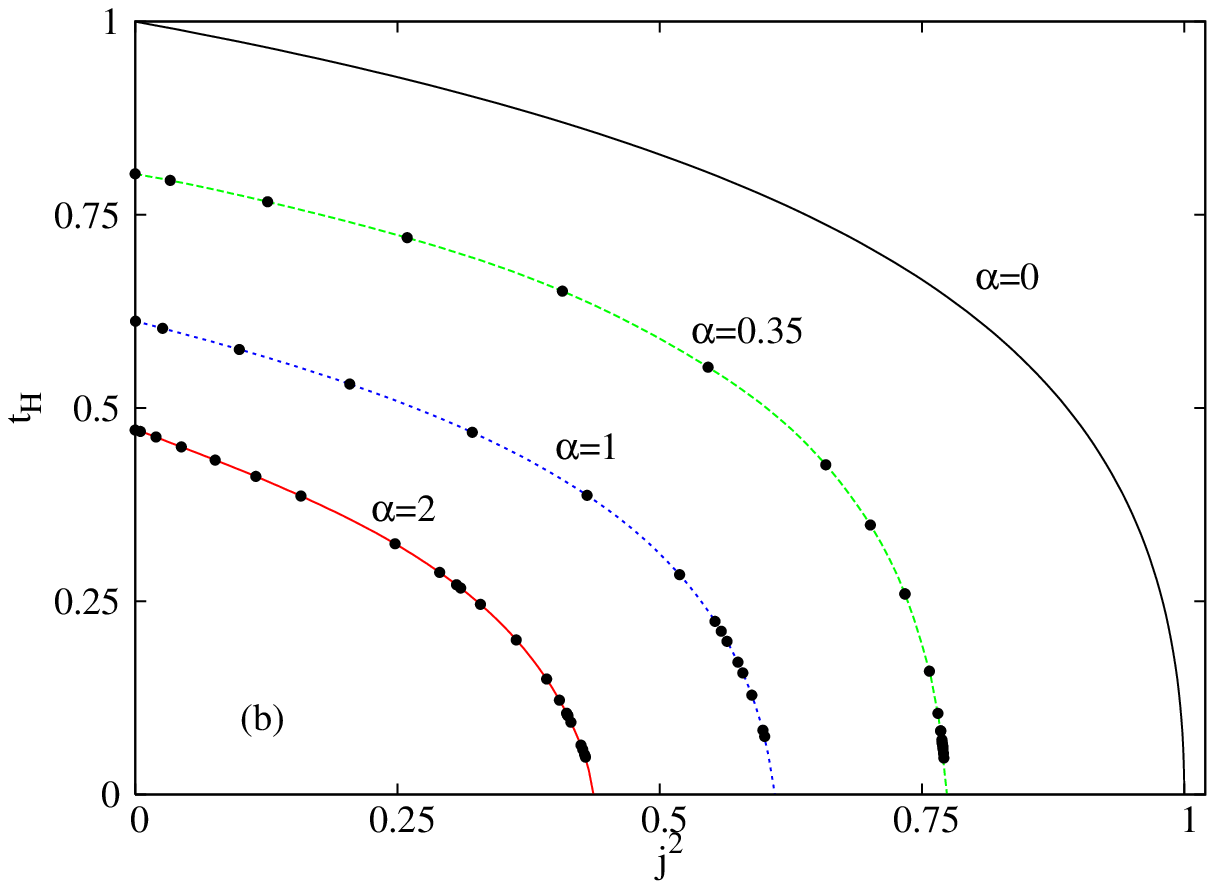}}	
\hss}
 \end{figure}
 %
\vspace*{-0.1cm}
 {\small \hspace*{3.cm}{\it  } }
\begin{figure}[ht]
\hbox to\linewidth{\hss%
	\resizebox{8cm}{6cm}{\includegraphics{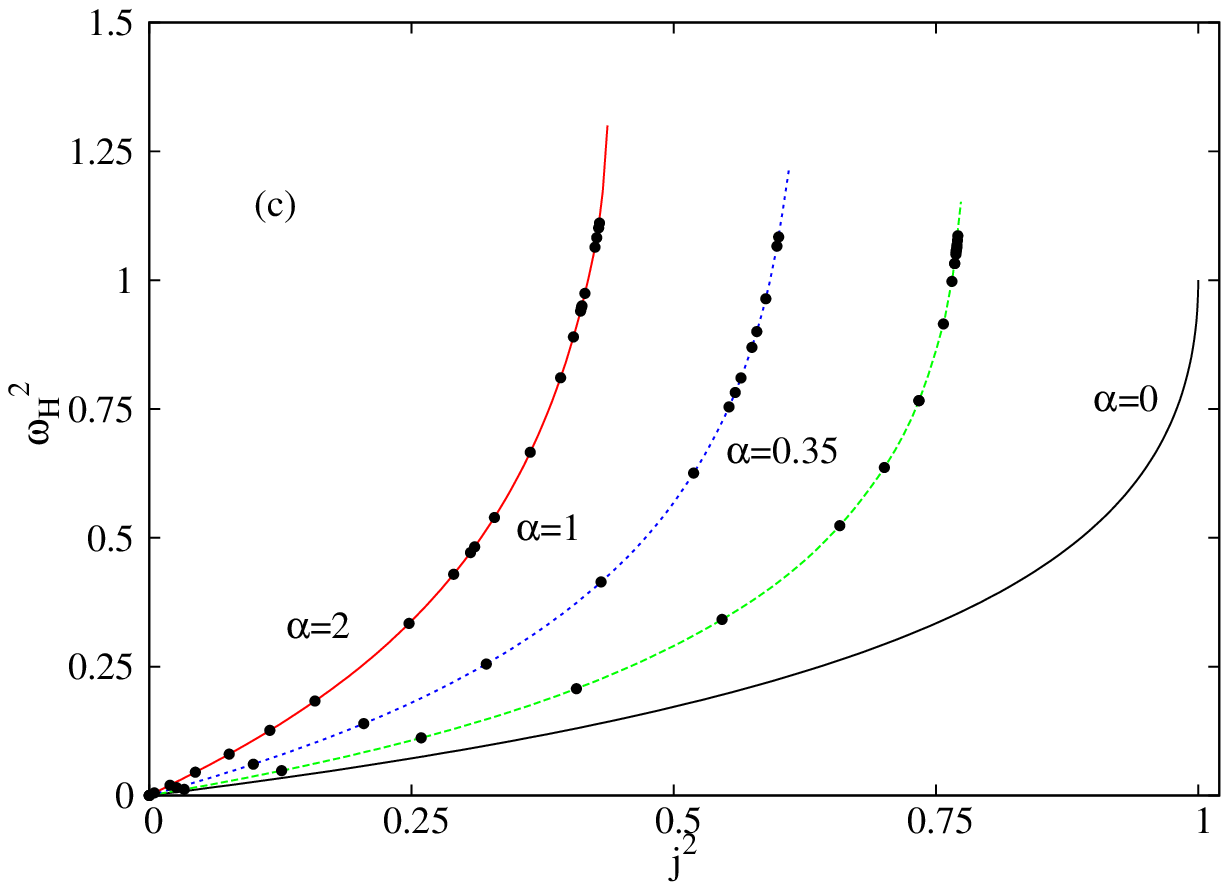}}
\hspace{10mm}%
        \resizebox{8cm}{6cm}{\includegraphics{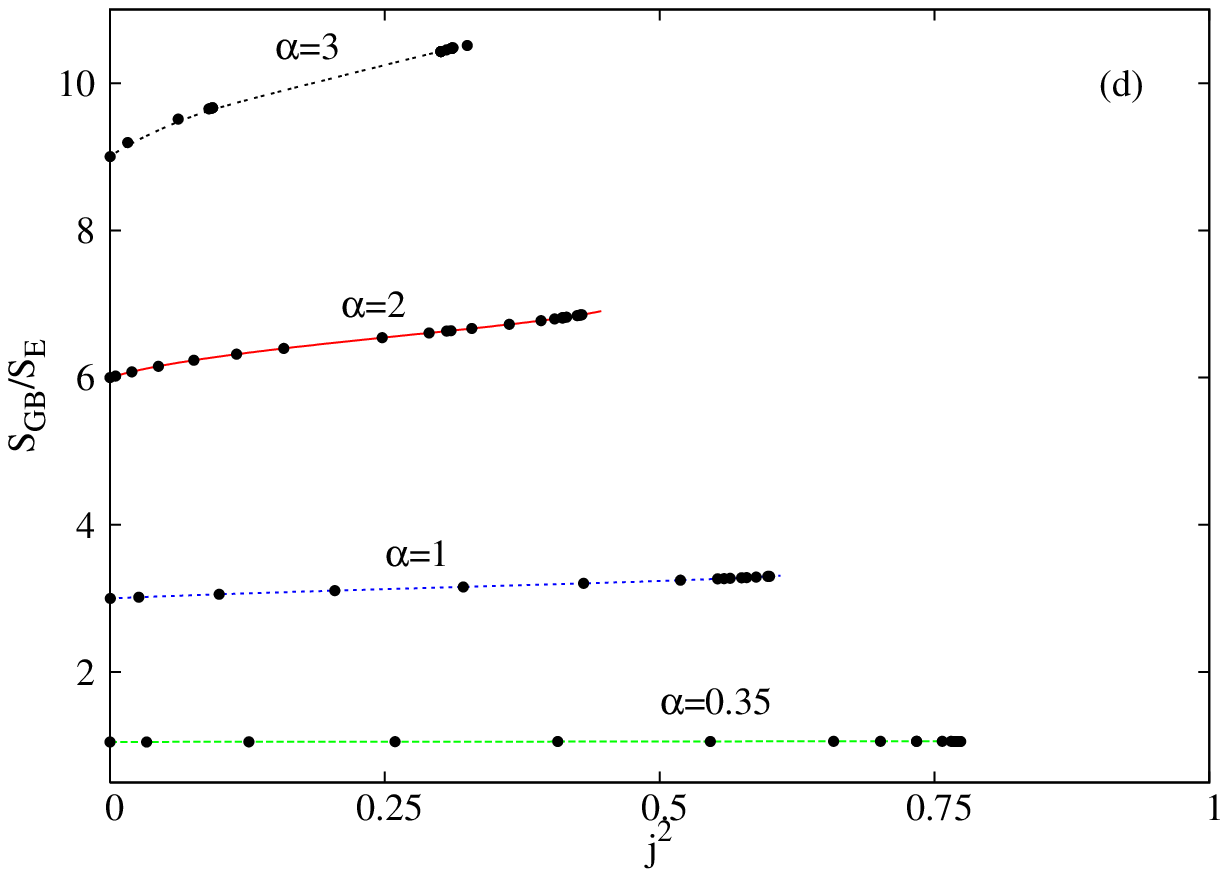}}	
\hss}
 \caption{\small 
The `reduced` dimensionless physical 
quantities 
(a) horizon area, (b) temperature, 
(c) horizon angular velocity, 
and (d) ratio between the Gauss-Bonnet
and the Einstein gravity contributions
to the total entropy are shown versus the dimensionless
squared (equal-magnitude) angular momenta for several
values of the Gauss-Bonnet coupling constant. 
The dots represent the data points. The spline-interpolated
curves have been extrapolated to the extremal endpoints.
  } 
\end{figure}
\newline
  out to be crucial. 

For $\Omega_H=0$, we know that $b'(r_H)= 2r_H/(r_H^2+\alpha)$, 
as can be seen from the explicit solution (\ref{SGB}). 
Then, increasing $\Omega_H$, our numerical results show
that
$b'(r_H)$ and $f'(r_H)$ both decrease monotonically.
This strongly suggests that they
reach the value zero in the limit $\Omega_H \to \Omega_H^{(c)}$.

 From these observations we conclude, 
that the families of rotating EGB black holes terminate
at extremal configurations. 
All relevant quantities remain finite as $\Omega_H \to \Omega_H^{(c)}$, 
while the Hawking temperature vanishes in the limit.
Moreover, we evaluated a number of curvature invariants 
($e.g.$~the scalar curvature $R$ 
and the Kretschmann scalar), and our extrapolations
showed that these stay finite everywhere in that limit.

This behaviour is thus analogous to that of MP black holes
in the extremal limit,
where $\Omega_H^{(c)}=1/\sqrt{2}r_H$.  
 We have found this picture for any value of $\alpha$
that we considered.
\newpage
 {\small \hspace*{3.cm}{\it  } }
\begin{figure}[t!]
\hbox to\linewidth{\hss%
	\resizebox{8cm}{6cm}{\includegraphics{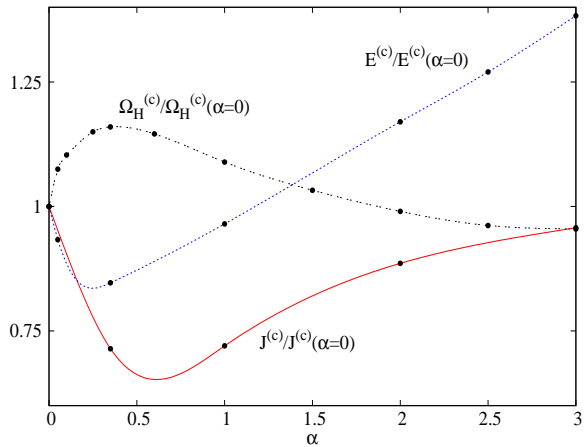}}
\hss}
\caption{\small 
The critical values of $\Omega_H$, $J$ and $E$
corresponding to extremal limiting solutions are shown as a function
of the Gauss-Bonnet coupling constant.
The dots represent the data points, the curves
are obtained by spline-interpolation.
}
\end{figure}
\\ 
 Therefore we conjecture that rotating black holes with equal-magnitude
 angular momenta exist for any value of
the GB coupling constant.  

To illustrate these aspects, we show in Figure 2
the behaviour of the `reduced' area of the horizon $a_H$, 
the `reduced' temperature $t_H$, the `reduced'
horizon angular velocity $\omega_H$, and the ratio $S_{GB}/S_E$
as functions of the squared `reduced' 
(equal-magnitude) angular 
momenta $j$ for several
values of the GB coupling constant $\alpha$.
These `reduced'  dimensionless  quantities are defined 
as follows 
\begin{eqnarray}
\label{def1}
a_H=\frac{3}{32}\sqrt{\frac{3}{2\pi G^3}}\frac{A_H}{E^{3/2}},
~~
t_H=4\sqrt{\frac{2\pi G}{3}}T_H\sqrt{E},
~~
\omega_H^2=\frac{8G}{3\pi}\Omega^2_H E,
~~
j^2=\frac{27\pi}{8G}\frac{J^2}{E^3}.
\end{eqnarray}
One can see that the pattern in Einstein 
gravity\footnote{The following relations hold for the solutions of
Einstein gravity
\begin{eqnarray}
 \label{defe}
a_H(j^2)=\frac{1}{2}(1+\sqrt{1-j^2})  ,
~~
t_H(j^2)= 2(1-\frac{1-\sqrt{1-j^2}}{j^2}),
~~
\omega_H^2(j^2)=\frac{2(1-\sqrt{1-j^2})}{j^2}-1.
\end{eqnarray} 
}
is recovered for all values of $\alpha$ considered.
A nonvanishing GB term decreases, however, the maximal values of the dimensionless quantities $j^2$,
$a_H$, and $t_H$, while it decreases the maximal value of $\omega_H$.

To  determine the domain of existence of the black holes 
with respect to the GB coupling constant $\alpha$,
we exhibit in Figure 3 the values of $\Omega_H$, $J$ and $E$
corresponding to critical solutions as functions of $\alpha$.
(Note, that these quantities are normalized there 
with respect to the corresponding ones for the MP solution).
The effect of the GB interaction is clearly perceptible on the diagram.
One should note that all these quantities 
have a nontrivial dependence on $\alpha$. 
 
 We conjecture that the domain of existence of 
rotating EGB black hole is the region {\it below} 
the (spline-interpolated) curve for $\Omega_H^{(c)}$ 
(or equivalently {\it below} the (spline-interpolated) line 
for the equal-magnitude angular momenta $J^{(c)}$). 
 
  \newpage
\vspace*{-0.1cm}
 {\small \hspace*{3.cm}{\it  } }
\begin{figure}[ht]
\hbox to\linewidth{\hss%
	\resizebox{8cm}{6cm}{\includegraphics{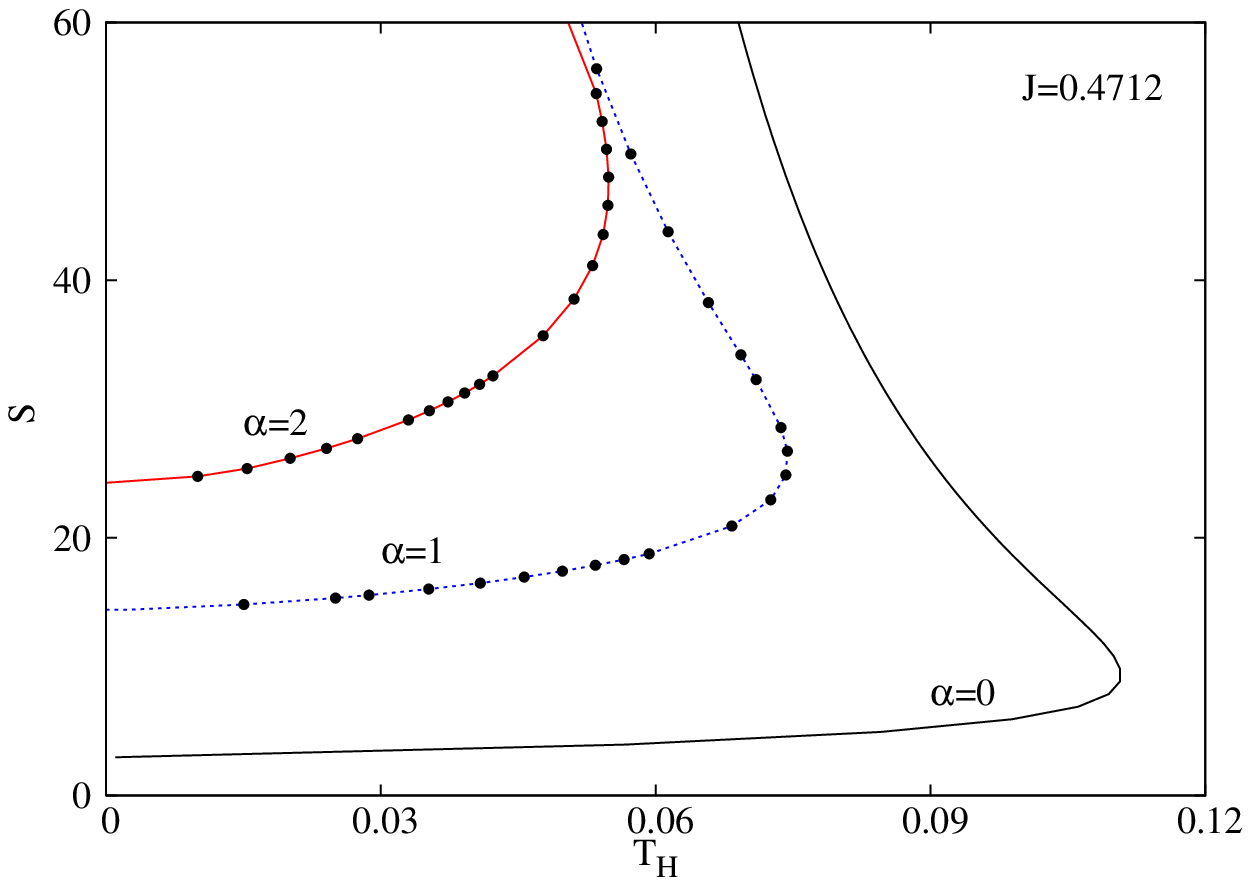}}
\hspace{10mm}%
        \resizebox{8cm}{6cm}{\includegraphics{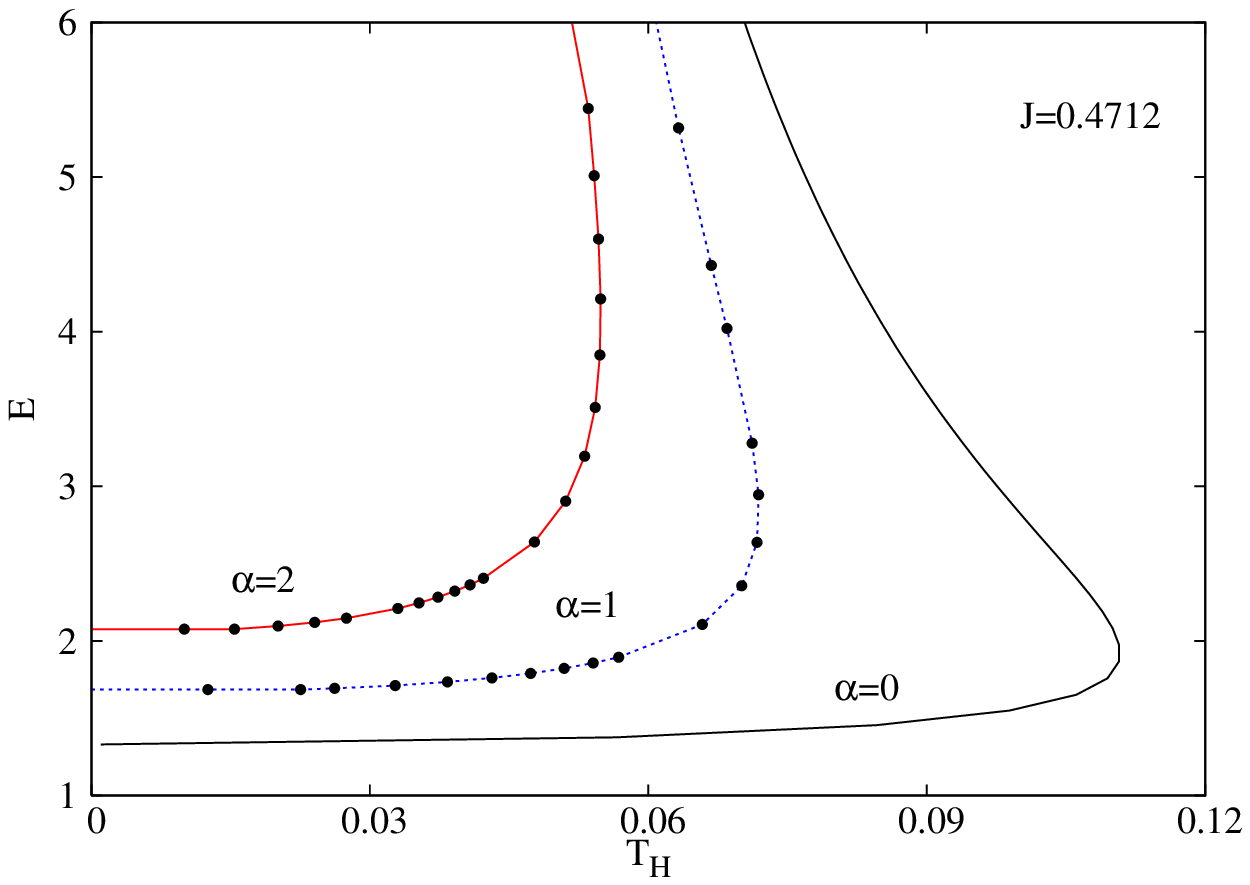}}	
\hss}
 \caption{\small 
The entropy $S$ and mass-energy $E$ are 
shown $vs.$ the Hawking temperature $T_H$ for fixed angular momenta $J$ 
 and several
values of the Gauss-Bonnet coupling constant (here and in Figures 5, 6 and 7 we set $G=1$). 
  } 
\end{figure}

 To control the quality of these results,
we have performed (a number of) these numerical calculations in two
different gauges, finding excellent agreement
for the physical parameters of the rotating EGB black holes. 
 Our systematic analysis was limited to $\alpha \in [0,3]$,
because the GB term is supposed to emerge
as a correction to the Einstein lagrangian; 
however, we also found families of solutions exhibiting the same
pattern for larger values of $\alpha$.

\subsubsection{Thermodynamical properties}
Considering the thermodynamics of these solutions,
the EGB black holes should satisfy the first law of thermodynamics 
\begin{eqnarray}
\label{firstlaw}
dE=T_HdS+2 \Omega_H dJ .
\end{eqnarray}
One may regard the parameters $S,~J$  as a complete set of extensive parameters
for the mass-energy $E(S,J)$ and define the intensive parameters
conjugate to them.
These quantities are the temperature and the angular velocities.
Also, for $\alpha\neq 0$, these solutions do not appear to satisfy any simple
Smarr relation\footnote{The Smarr relation in Einstein gravity is 
$E=3(T_H S+2\Omega_H J)/2$. The recent work \cite{Kastor:2010gq} proposes
a Smarr relation for static black holes in Lovelock gravity.
It would be interesting to extend the formalism in  \cite{Kastor:2010gq} to the
case of spinning solutions. }.

In the absence of an exact solution, 
we attempt here to analyze  the thermodynamic stability of the rotating EGB solutions
based on the  available numerical data.
Since we did not yet explore the full parameter space of solutions,
the results below are only partial, and one cannot exclude  
the existence of new features outside the explored domain. 

It is known that different thermodynamic ensembles are not
exactly equivalent and may not lead to the same conclusions since they correspond to different
physical situations 
\newpage
\vspace*{-0.1cm}
 {\small \hspace*{3.cm}{\it  } }
\begin{figure}[ht]
\hbox to\linewidth{\hss%
	\resizebox{8cm}{6cm}{\includegraphics{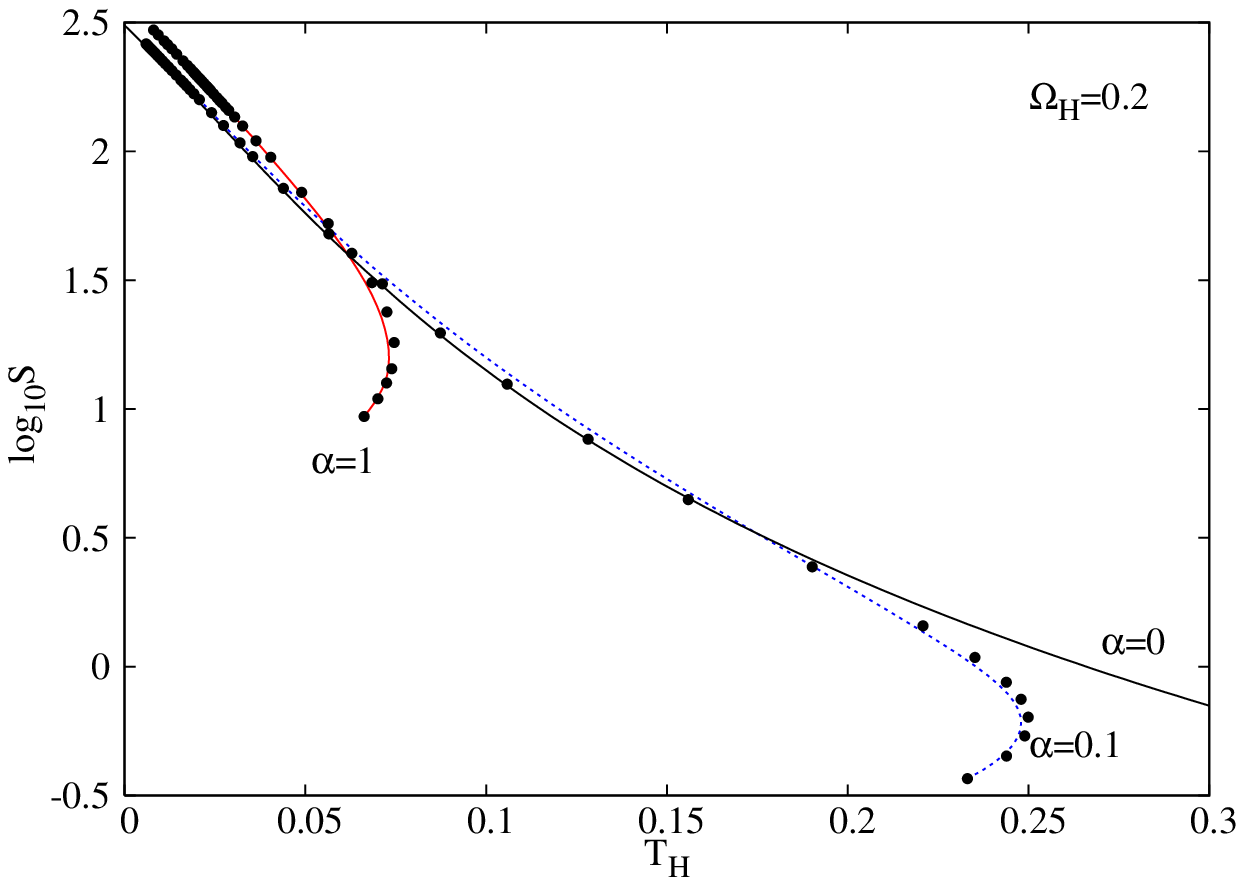}}
\hspace{10mm}%
        \resizebox{8cm}{6cm}{\includegraphics{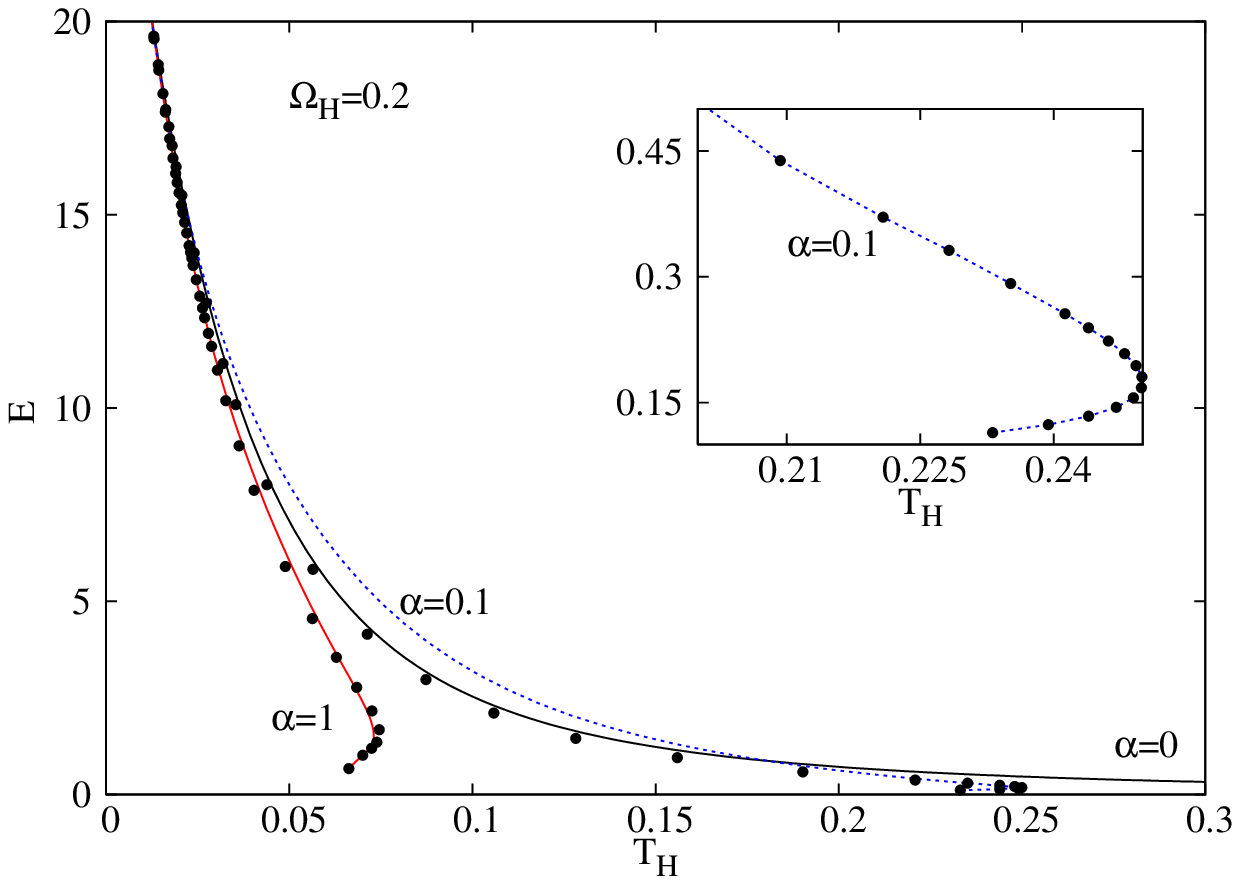}}	
\hss}
 \caption{\small 
The entropy $S$ and mass-energy $E$ are 
shown $vs.$ the Hawking temperature $T_H$ for fixed horizon angular velocity $\Omega_H$ 
 and several
values of the Gauss-Bonnet coupling constant. 
  } 
\end{figure}
\\
(see $e.g.$ \cite{Monteiro:2009tc}, \cite{Astefanesei:2009wi}, \cite{Astefanesei:2010bm}
for recent work on the thermodynamics
of spinning black objects in Einstein gravity).

In the canonical ensemble, we study black holes holding the temperature
$T_H$ and the angular momenta $J$  fixed,
the associated thermodynamic potential being the Helmholz free energy $F[T_H,J]=E-T_HS$.
In this case, the numerical analysis for several values of $\alpha$ indicates that the qualitative
thermodynamical features of the  MP solutions are also shared by their EGB
generalizations. 
The configurations near extremality ($i.e.$ with a small enough $T_H$) 
are thermally stable in a canonical ensemble since $C_J=T_H(\partial S/\partial T_H)|_{J} >0$.
However, there is also a branch of large black holes whose entropy is a decreasing function of $T_H$.
At the critical point, the specific heat goes through an infinite discontinuity, 
and a phase transition takes place.
This is the picture one finds for the MP solution in Einstein gravity, in which case the
specific heat changes the sign for 
$T_H^{c}\simeq 0.087396/(GJ)^{1/3}$.
Interestingly, 
for a given $J$, the effect of the GB term is 
to decrease the critical value of the Hawking temperature.
These features are shown in Figure 4 where we exhibit both the entropy and the mass-energy
of the solutions as functions of $T_H$
for an arbitrary fixed value of $J$ and two nonzero values of $\alpha$
as well as for $\alpha=0$.

In the grand canonical ensemble, on the other hand, we keep the temperature and the horizon
angular
velocity fixed. In this case the thermodynamics is obtained from the Gibbs
potential $G[T_H,\Omega_H]=E-T_HS-2\Omega_H J=I_{cl}/\beta$.
The first quantity of interest here is the specific heat at constant horizon angular velocity 
$C_\Omega=T_H(\partial S/\partial T_H)|_{\Omega_H} >0$.
A straightforward computation  shows that for the
Myers-Perry black hole this is a negative quantity  
\begin{eqnarray}
\label{sa1}
C_\Omega=-\frac{1}{4 \pi G T_H^4}\frac{\sqrt{2+x^2-2\sqrt{1+x^2}}(2+2x^2+\sqrt{1+x^2})}
{x^2(1+x^2)^{3/2}(1+x^2+\sqrt{1+x^2})}<0,
 ~~ {\rm~with~~}x=\frac{\Omega_H}{\pi T_H}.
\end{eqnarray}
\newpage
 {\small \hspace*{3.cm}{\it  } }
\begin{figure}[t!]
\hbox to\linewidth{\hss%
	\resizebox{8cm}{6cm}{\includegraphics{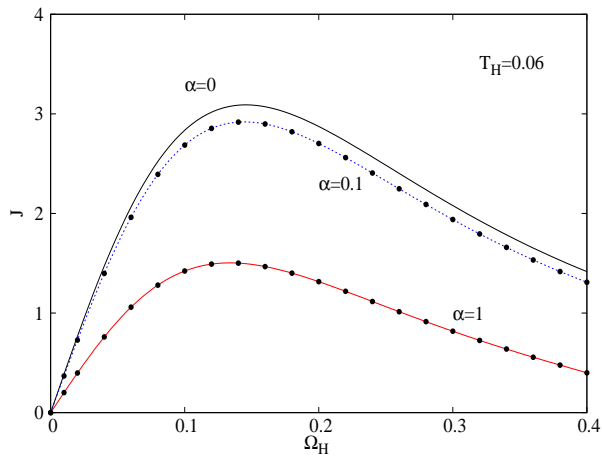}}
\hss}
\caption{\small 
The angular momentum $J$ is shown as a function of
the horizon angular velocity $\Omega_H$ for a fixed Hawking 
temperature and several values
of the Gauss-Bonnet coupling constant.}
\end{figure}
\\
However, not completely unexpected\footnote{We recall the existence of a
 branch of small static
EGB black holes that is thermodynamically stable \cite{Myers:1988ze}.},
our numerical results show the existence 
a branch of rotating black holes with $C_\Omega>0$
for any value of $\alpha>0$ considered.

To illustrate this feature, we show in Figure 5 the entropy and the mass-energy
of solutions as functions of the Hawking temperature for several values of the GB coupling constant
and an arbitrary fixed value of the horizon angular velocity (for better visualization, we used a 
logarithmic scale for $S$).
We observe that the critical temperature where $C_\Omega$
changes sign decreases with increasing $\alpha$,
while the Einstein gravity picture is recovered for large black holes,
where $C_\Omega<0$.

Another response function of interest is the isothermal permittivity
$\epsilon_{T_H}=(\partial J/\partial \Omega_H)|_{T_H}$.
The Einstein gravity result is
\begin{eqnarray}
\label{sa2}
\epsilon_{T_H}=  \frac{V_3}{16 \pi^5 G T_H^4}
\frac{ 5+6x^2-\sqrt{1+x^2}(5+3x^2)}
{x^2(1+x^2)^{3/2}(1+\sqrt{1+x^2})^2},
~~{\rm~~ where~~}x=\frac{\Omega_H}{\pi T_H},
\end{eqnarray}
and thus $\epsilon_{T_H}>0$ for ${\Omega_H}/{\pi T_H}\lesssim 0.77272$.
Our results for $\alpha=0.1$ and $\alpha=1$ show that 
$\epsilon_{T_H}$ changes sign also for EGB solutions and that,
for a given $T_H$, the solutions with a small $\Omega_H$ have 
a positive isothermal permittivity (see Figure 6).
Thermodynamic stability in the grand-canonical ensemble
requires that the specific heat at constant
angular momentum $C_J$, the isothermal permittivity  $\epsilon_{T_H}$ 
as well as the specific heat at constant horizon angular velocity $C_\Omega$
are positive.
Although a systematic study of the domain of thermodynamic stability 
of the EGB solutions is beyond the purposes of this paper,
our results show the existence of solutions which fulfill these conditions.
The stability conditions hold for a set
of rotating solutions 
emerging from small static black holes (that have positive specific heat).

We expect that this result is relevant for the 
properties of the corresponding $d=6$
rotating  black string solutions in EGB theory (note that such solutions
cannot be found by simply uplifting the black holes in this paper).
For example, the results in  \cite{Kleihaus:2007dg} indicate that
the Gregory-Laflamme  instability \cite{Gregory:1993vy} persists up to extremality for all $d=6$
Einstein gravity black strings with equal magnitude angular momenta.
Based on the Gubser-Mitra conjecture \cite{Gubser:2000ec}  that correlates the dynamical and the thermodynamical stability,
we conjecture 
that in a certain region of parameter space there are spinning EGB solutions which do not
possess a Gregory-Laflamme instability.
Also, not completely unrelated, it would be interesting to study the thermodynamic stability of these solutions
from the perturbative corrections to the gravitational partition function
(see \cite{Monteiro:2009ke} for a recent similar study of the Kerr-AdS black holes).

We conclude that the presence
of a GB term in the lagrangian affects the thermodynamical
properties of the solutions and allows for
 thermodynamically stable solutions,
both in the macrocanonical and the
canonical ensemble.

\subsection{Extremal solutions: 
near horizon geometry and the entropy function}

Returning to the issue of the extremal solutions,
we recall that the numerical integration in the neighbourhood 
of extremal black holes is very difficult, 
as noted also in other cases.
The near horizon expansion of extremal configurations  is still given by 
(\ref{c1}), with $f_1=b_1=0$, and thus they satisfy a different 
set of boundary conditions at $r=r_H$ than the nonextremal configurations obtained  in this work.
Therefore finding such solutions explicitly 
is beyond the scope of this paper\footnote{Note, however, that we could construct with
relatively good accuracy {\it near-extremal} black holes.}. 
Most likely, this should require a different parametrization for the metric,
better suited to the extremal case.

However, we argue that the existence of the EGB generalizations 
of the extremal MP black holes for any $\alpha>0$
is strongly suggested by the existence of an exact solution
describing a rotating squashed $AdS_2\times S^3$ spacetime. 
This solution would describe the neighbourhood of the event horizon of an
 extremal black hole.
(The far field expression of the extremal solution is still given by (\ref{inf1}), 
with a single essential parameter in the expansion there.)

Therefore we consider the following metric form (see the generic ansatz (\ref{metric2})) in corotating coordinates
\begin{eqnarray}
\label{at1}
ds^2=v_1(\frac{dr^2}{r^2}-r^2 dt^2)+\frac{v_2}{4}(\sigma_1^2+\sigma_2^2)
+\frac{v_2v_3}{4}(\sigma_3+2 k r dt)^2
\end{eqnarray}
($i.e.$ for $b(r)=v_1 r^2$, $f(r)=r^2/v_1$, $g(r)=v_2$,
$h(r)=v_2 v_3$, $w(r)=k r$ within the parametrization in this paper),
such that the horizon is located\footnote{This position of the horizon can always be obtained by taking
$r\to r-r_H$.} at $r=0$.
This geometry describes a fibration of $AdS_2$ over the homogeneously squashed $S^3$ with symmetry
group $SO(2, 1)\times SU(2)\times U(1)$ \cite{Kunduri:2007qy}.

The parameters $v_i,k$ satisfy a set of algebraic relations which result from 
the EGB equations.
In what follows we choose to determine them by using the  
formalism proposed in \cite{Astefanesei:2006dd}, thus by
 extremizing 
an entropy function.
This allows us also to compute the entropy of these black holes 
and to show that the solutions exhibit attractor behaviour.
 
Therefore let us denote by $f(k,\vec v)$ the lagrangian density $\sqrt{-g} {\cal L}$
evaluated for the near horizon geometry (\ref{at1})
and integrated over the angular coordinates,
\begin{eqnarray}
\label{at2}
f(k,\vec v)&=&\int d\bar \theta d \phi d\psi \sqrt{-g} {\cal L}
=\frac{1}{16 \pi G} \int d\bar \theta d \phi d\psi \sqrt{-g}  (R +\frac{\alpha}{4}L_{GB})  .
\end{eqnarray}
The metric field equations in the near horizon geometry (\ref{at1}) 
 now correspond to
$\frac{\partial f}{\partial k}=J,~~\frac{\partial f}{\partial v_i}=0,$
with $J$ the angular momenta of the solutions.

Then,
following  \cite{Astefanesei:2006dd}, we define the entropy function by taking the Legendre transform of the above integral
with respect to the parameter $k$,
\begin{eqnarray}
\label{at4}
{\cal E}(J,k,\vec v)=2 \pi (J k-f(k,\vec v)).
\end{eqnarray}
It follows as a consequence of the equations of motion that 
the constants $k,\vec v$
are solutions of the equations
  \begin{eqnarray}
 \label{at32}
 \frac{\partial {\cal E}}{\partial k}=0,~~\frac{\partial {\cal E}}{\partial v_i}=0. 
  \end{eqnarray}
Then, the entropy associated with the  black hole is given by
$S_{extremal}={\cal E}(J,k,\vec v)$
evaluated at the extremum (\ref{at32}).
Further details on this formalism and explicit examples are given $e.g.$ in
\cite{Cai:2007uw}, \cite{Goldstein:2007km}, \cite{Astefanesei:2007bf}.

For the metric ansatz (\ref{at1}), a straightforward calculation gives
\begin{eqnarray}
\label{at21}
{\cal E}(J,k,\vec v)=
2\pi \bigg [
Jk -\frac{\pi \sqrt{v_2 v_3}}{16 G v_1}
\left(
k^2 v_2^2 v_3-4 v_1^2 (v_3-4)-4 v_1 v_2
-\alpha (k^2 v_2 v_3(3v_3-4)-4v_1(v_3-4)
\right)
\bigg ],~~{~~~}
\end{eqnarray}
such that 
the explicit form of the  equations (\ref{at32}) is
\begin{eqnarray}
\label{at5v}
&&\frac{\partial {\cal E}}{\partial v_1}=0 
\Rightarrow 
 -16 v_1^2+4 v_1^2 v_3+k^2 v_2^2 v_3+
\alpha k^2 v_2v_3(4-3v_3)=0,
\\
\nonumber
&&
\frac{\partial {\cal E}}{\partial v_2}=0  \Rightarrow 
-16 v_1^2+12 v_1 v_2+4 v_1^2v_3-5k^2 v_2^2 v_3+
\alpha(-4v_1(v_3-4)+3k^2v_2 v_3(3v_3-4)) =0,
\\
\nonumber
&&
\frac{\partial {\cal E}}{\partial v_3}=0  \Rightarrow 
-16 v_1^2+4 v_1 v_2+12 v_1^2v_3-3k^2 v_2^2 v_3+
\alpha(-4v_1(3v_3-4)+3k^2v_2 v_3(5v_3-4)) =0,
\end{eqnarray}
and
\begin{eqnarray}
\label{at5}
\frac{\partial {\cal E}}{\partial k}=0 
\Rightarrow 
J=\frac{\pi k (v_2 v_3)^{3/2}}{8 G v_1}
( v_2+\alpha (4-3v_3) ).
 \end{eqnarray}
 In Einstein gravity the solution has a simple form in terms of $J$
 \begin{eqnarray}
\label{at6}
v_1=\frac{1}{2}
\left(\frac{GJ}{2\pi}\right)^{2/3},~~
v_2= 
\left(\frac{\sqrt{2} GJ}{ \pi}\right)^{2/3},~~v_3=2,~~k=\frac{1}{2},
\end{eqnarray}
and thus the known result ${\cal E}=S_{extremal}=\pi J$ is recovered.

In the limit of small $\alpha$,
one can treat
the GB term as a perturbation. 
In second order in $\alpha$,
one finds the following solution of the system (\ref{at5v}), (\ref{at5}) in terms of the global charge $J$
 \begin{eqnarray}
\label{sat1}
&&v_1=\frac{1}{2}
\left(\frac{GJ}{2\pi}\right)^{2/3}
-\frac{\alpha}{6}
+\left(\frac{\pi}{\sqrt{2}G J} \right)^{3/2}\frac{\alpha^2}{9},~~
v_2= 
\left(\frac{\sqrt{2} GJ}{ \pi}\right)^{2/3}
+\frac{4}{3}\alpha+\left(\frac{2\pi}{ G J} \right)^{3/2}\frac{2\alpha^2}{9},~~~~{~~~}
\\
\nonumber
&&v_3=2
-2\left(\frac{ 2 \pi }{GJ}\right)^{2/3}\alpha
-\left(\frac{ \pi }{GJ}\right)^{4/3}\frac{ 2^{1/3}16\alpha^2}{3},
~~k=\frac{1}{2}+\left(\frac{  \pi }{\sqrt{2}GJ}\right)^{2/3}\frac{\alpha}{2}
+\left(\frac{ \pi }{GJ}\right)^{4/3}\frac{ \alpha^2}{2^{2/3}12}~,
\end{eqnarray}
while
 \begin{eqnarray}
\label{sat2}
{\cal E}=\pi J+3\left(\frac{ J\pi^5}{ 2G^2}\right)^{1/3}\alpha
-\left(\frac{  \pi^7}{ 4G^4 J}\right)^{1/3}\frac{\alpha^2}{2}.
\end{eqnarray}
Therefore the approximate solutions exhibit a complicated behaviour in terms of $\alpha$ and $J$ and 
a perturbative approach may be misleading. 
 
Unfortunately, for a nonzero GB term in the action, it seems that the only possibility 
is to express the nonperturbative solution of the system (\ref{at5v}), (\ref{at5})
in terms of the relative squashing parameter $v_3$, with 
 \begin{eqnarray}
\label{at7}
&&v_1=-\frac{(v_2+(4-3v_3)\alpha)(3v_2-(v_3-4)\alpha)}{2(v_3-4)(3v_2+(8-6v_3)\alpha) },
\\
\label{at71}
&&J=\frac{\pi v_2 v_3}{G}\sqrt{(4-v_3)(v_2+\alpha (4-3v_3))},~~{~~}
\end{eqnarray}
and 
 \begin{eqnarray}
\label{at8}
k=\frac{8 G J v_1}{\pi (v_2 v_3)^{3/2}(v_2+(4-3v_3)\alpha)},
\end{eqnarray}
while the radius of the round $S^2$ sphere in the line element (\ref{at1}) 
is given by\footnote{Here we restrict to the physical solution which recovers the general
relativity limit as $\alpha\to 0$.}
 \begin{eqnarray}
\label{at9}
v_2=\frac{\alpha}{v_3-2} 
\left(
2v_3^2-7v_3+4
-\sqrt{5v_3^4-34 v_3^3+73 v_3^2-56 v_3+16}
\right).
\end{eqnarray}
Also, we notice that the relations (\ref{at7})-(\ref{at9})
are invariant under the scaling
\begin{eqnarray}
\nonumber
  v_1 \to \lambda v_1,~
v_2 \to \lambda v_2, ~
{\cal E} \to \lambda^{3/2} ~
{\cal E} , ~
J \to \lambda^{3/2} J,~
k \to k,~{\rm and}~\alpha \to \lambda \alpha,
\end{eqnarray}
which shows that the solutions exist for any $\alpha\geq 0$. 

Inserting these expressions into Eq.~(\ref{at21})
 we obtain for the entropy function of the
extremal black hole:
 \begin{eqnarray}
\label{at10}
{\cal E}=S_{extremal}=  \frac{\pi^2}{2G}\sqrt{v_2 v_3}\left(v_2-(v_3-4)\alpha \right),
\end{eqnarray}
 (with $v_2(v_3)$ as implied by (\ref{at9})).

\newpage
\vspace*{-0.1cm}
 {\small \hspace*{3.cm}{\it  } }
\begin{figure}[ht]
\hbox to\linewidth{\hss%
	\resizebox{8cm}{6cm}{\includegraphics{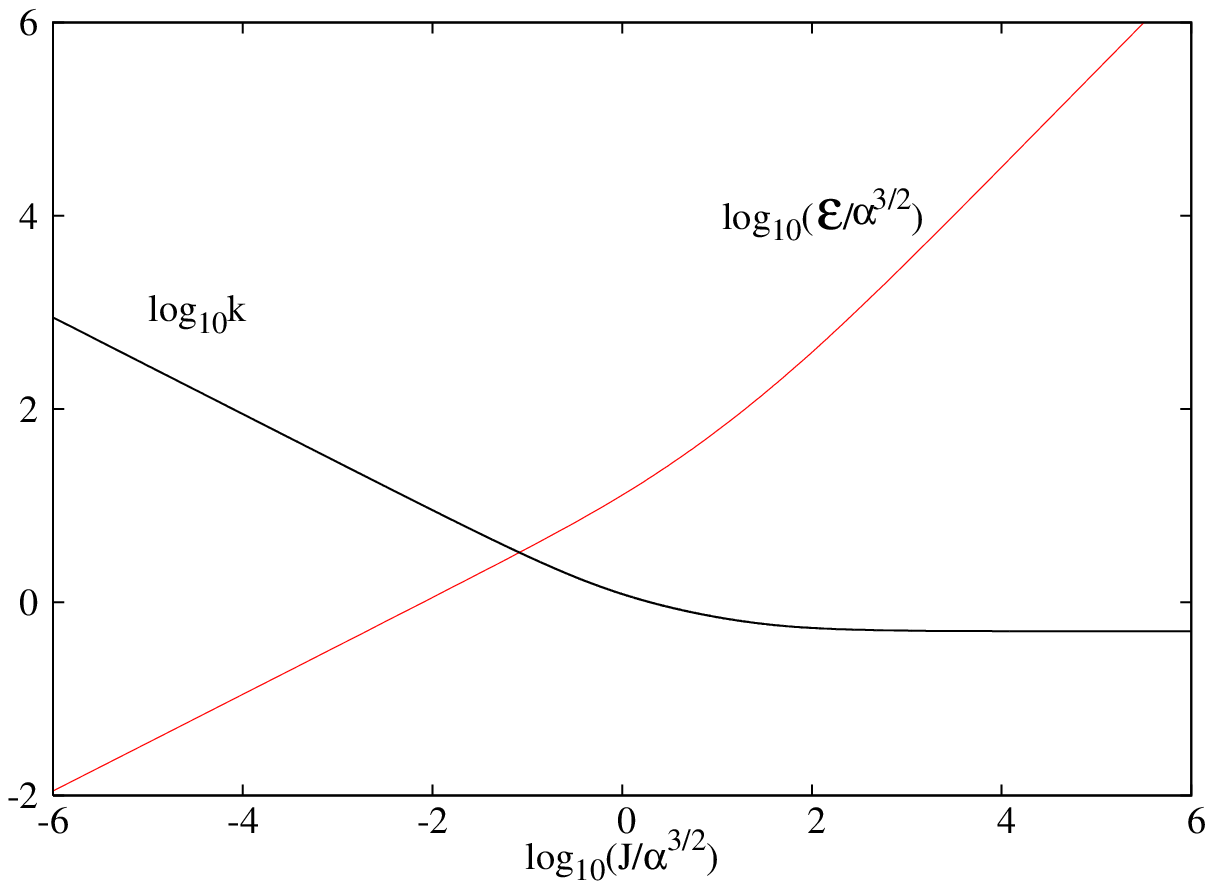}}
\hspace{10mm}%
        \resizebox{8cm}{6cm}{\includegraphics{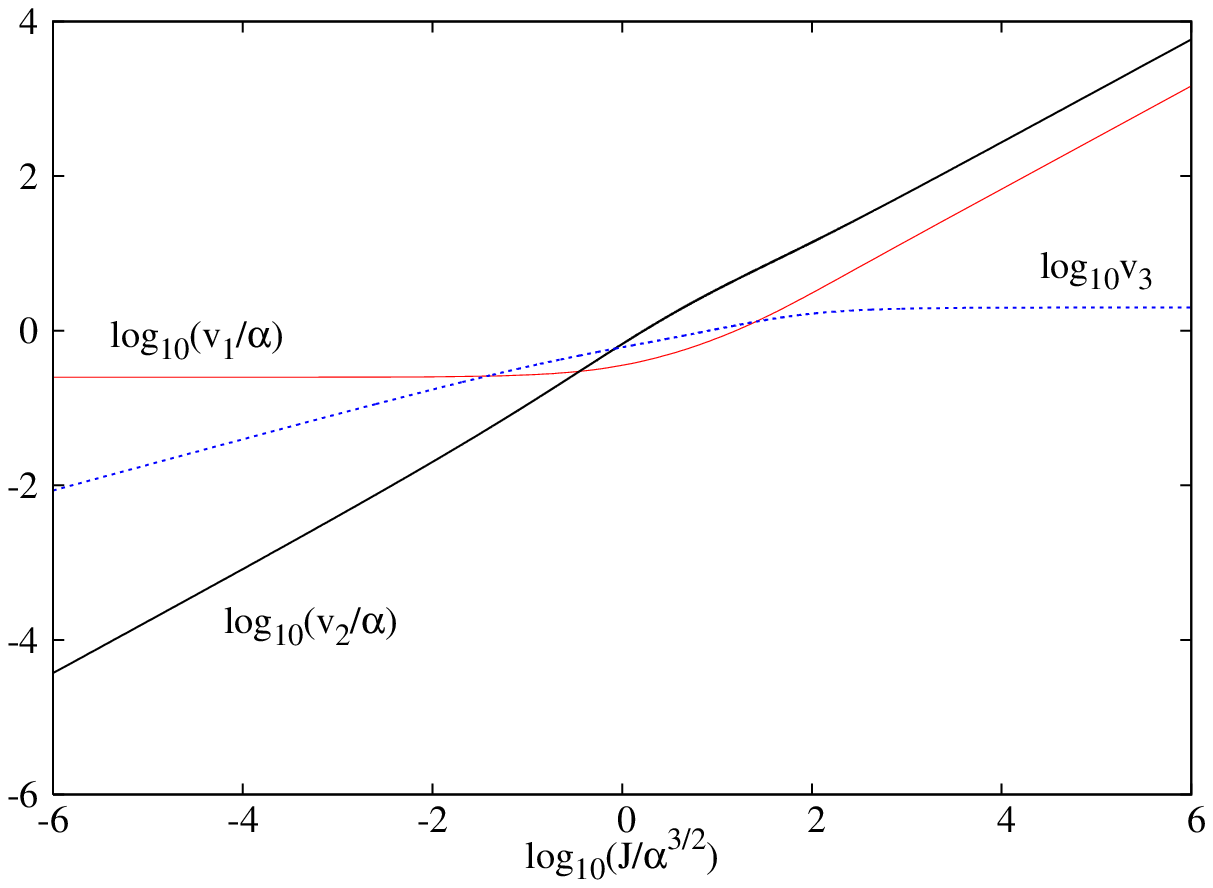}}	
\hss}
 \caption{\small 
The dimensionless  quantities ${\cal E}/\alpha^{3/2}$, $k$, $v_1/\alpha$, 
$v_2/\alpha$ and $v_3$ are shown as functions of  $J/\alpha^{3/2}$  
(for a better visualisation, we employ here logarithmic scales).
  } 
\end{figure}
 
As a check, we note that the result (\ref{at10}) 
agrees with Wald's form (\ref{S-Noether}) 
evaluated for the near horizon geometry (\ref{at1}).
We also note that, in principle, ${\cal E}$ can be expressed in terms of the conserved
charge $J$ by inverting relation (\ref{at71}).

We conclude that a nonzero $\alpha$ may 
substantially affect
the near horizon geometry of an extremal black hole.
For example, the allowed range\footnote{This results from the physical condition that $v_i,J^2$
are strictly positive quantities.} for the 
relative squashing parameter $v_3$ is
$0<v_3\leq 2$ ($i.e.$ the GB term reduces the relative squashing of the solutions),  with 
 \begin{eqnarray}
\label{at11}
\nonumber
&v_1=\frac{\alpha}{4}+\dots,~v_2=\frac{\alpha v_3^2}{2}+\dots,~J=\frac{\pi \alpha^{3/2}v_3^3}{2G}+\dots,
~
k=\frac{1}{\sqrt{2}v_3^{3/2}}+\dots,~{\cal E}=\frac{\sqrt{2}\pi^2 }{G} (\alpha v_3)^{3/2} +\dots, 
\end{eqnarray}
as $v_3 \to 0$, and
 \begin{eqnarray}
\label{at12}
\nonumber
& v_1=\frac{\alpha}{2-v_3}+..,~v_2=\frac{4 \alpha }{2-v_3}+..,
~J=\frac{4\sqrt{2}\pi  }{ G }\left({\frac{\alpha}{2-v_3}}\right)^{3/2}+.., 
 ~k=\frac{1}{2}+..,~
 {\cal E}=\frac{4\sqrt{2}\pi^2 }{G} \left({\frac{\alpha}{2-v_3}}\right)^{3/2} +.., 
\end{eqnarray}
as $v_3\to 2$.
In Figure 7 
we exhibit a number of relevant dimensionless quantities as functions of 
the scaled angular momenta $J/\alpha^{3/2}$ (where for better visualisation,
we use logarithmic scales).
We observe that
the ratio ${\cal E}/J$ is no longer constant in EGB
gravity. 

However, finding 
local solutions in the vicinity of the horizon
does not guarantee the existence of global asymptotically flat solutions.
Further progress in this direction seems to require an explicit construction
of the bulk extremal black hole solutions. 
For example, this would also allow to construct the $E(J)$ diagram for such configurations.

{
We close this Section by remarking that the study of these
$AdS_2\times S^3$ solutions in EGB theory
is interesting in yet another context. 
In ref. \cite{Guica:2008mu} it has been 
proposed that the near horizon geometry of an extremal Kerr
black hole is holographically dual to a 2-dimensional chiral conformal
field theory (CFT).
This correspondence has been extended to various other examples of extremal
spinning black holes in
$d\geq 4$ dimensions, including configurations with matter fields.
These studies are based on the
universality character of the near horizon geometry of extremal black holes.
It would be interesting to consider also the case
of  such solutions with GB corrections,
the configurations in this work being perhaps the simplest 
relevant example.
In a Kerr/CFT context,
 the entropy formula (\ref{at10}) should  be recovered
by computing the central charge of a certain two-dimensional
conformal algebra\footnote{Note that one can associate a 
temperature $T=1/2\pi k$ to the near horizon geometry (\ref{at1}).
As can be seen from Figure 7, in EGB theory
this temperature is no longer constant $T=1/\pi$,
presenting a a nontrival dependence on the dimensionless ratio $G J/\alpha^{3/2}$.}.
}

We hope to return to the study of extremal black holes in EGB theory 
in future work.

\section{Further remarks }

The main purpose of this paper was to
present evidence for the existence of 
rotating black holes in $d=4+1$ EGB theory.
Representing generalizations
of a particular class of MP black holes,
the considered configurations possess
a regular horizon of spherical topology and 
two equal-magnitude angular momenta.
  Our results indicate that the inclusion of a GB term in the action 
does not affect most of the qualitative features of the solutions.  
However,  the presence of a GB term in the action
has a tendency to stabilize the rotating black holes,
leading to  
a branch of solutions with a
positive
specific heat at constant angular velocity at the horizon
which does not exist for the MP solutions.

Also, analogous to the case of Einstein gravity,  
when the horizon angular velocity is increased,
the black hole solutions reach a limiting extremal black hole 
with a regular horizon.
 Although we did not attempt to construct
these extremal black holes, 
we gave further support for their existence
in Section 4.2
by finding an exact $AdS_2\times S^3$ rotating solution
in EGB theory. 
This solution would describe the neighbourhood of the event horizon of an
 extremal EGB black hole.

A natural question that arose  during our study was  how to determine the various conserved
quantities and the total action of the 
solutions. One particularly powerful
approach to this problem is given by the counterterms method.
In Section 3 we showed how to generalize the  Einstein gravity 
counterterms in \cite{Mann:1999pc,Lau:1999dp,Kraus:1999di}
by including  the effects of a GB term in the bulk action.
Although the counterterm method gives results that are equivalent to those obtained using
the background subtraction method, we employed it here because it appears to
be a more general technique than background subtraction. Moreover, it is interesting
to explore the range of problems to which it applies, in particular for configurations with 
higher order curvature corrections in the gravity action.

The solutions obtained in this paper 
may provide a fertile ground for the further study of
rotating configurations in EGB theory.
For example, their generalization to include the
effects of an electromagnetic field is straightforward.
Also, in principle, by using the same techniques,
there should be no difficulty to construct similar solutions 
in $d=2N+1$ dimensions
with $N>2$ equal-magnitude angular momenta.
 
Another interesting direction to consider in future work 
consists in finding the EGB generalizations of the $d=5$ MP black holes 
with nonequal angular momenta, in particular 
the case with rotation in a single plane.
Concerning the latter case, 
it would be interesting to see how the GB term would affect 
the properties of the solutions close to the extremal limit.
Different from the solution (\ref{MP}), 
in this limit the Einstein gravity solution corresponds 
to a naked singularity.
One might speculate that the higher derivative terms in the action 
might smoothen this singularity 
and lead to a physically reasonable solution.
These aspects together with the investigation
of the effects of the GB term on balanced black rings
and thus the phase diagram 
are presently under investigation.
\\
\\
{\bf\large Acknowledgements} 
\\
YB is grateful to the Belgian FNRS for financial support.
BK gratefully acknowledges support by the DFG.
The  work of ER
is carried out in the framework of Science Foundation Ireland (SFI) project RFP07-330PHY.

 
\end{document}